\documentclass[sigconf]{acmart}
\usepackage{multirow}
\usepackage{makecell}
\usepackage{subfig}
\usepackage{bbding}

\copyrightyear{2022}
\acmYear{2022}
\setcopyright{acmcopyright}
\acmConference[SIGIR '22]{Proceedings of the 45th International ACM SIGIR Conference on Research and Development in Information Retrieval}{July 11--15, 2022}{Madrid, Spain.}

\acmBooktitle{Proceedings of the 45th International ACM SIGIR Conference on Research and Development in Information Retrieval (SIGIR '22), July 11--15, 2022, Madrid, Spain}
\acmPrice{15.00}
\acmISBN{978-1-4503-8732-3/22/07}
\acmDOI{10.1145/3477495.3531970}

\begin{document}
\fancyhead{}

\title{Enhancing CTR Prediction with Context-Aware Feature Representation Learning} 

\author{Fangye Wang}
\authornote{
Also Shanghai Key Laboratory of Data Science, Fudan University, Shanghai, China.
}
\orcid{0000-0001-7216-1688}
\affiliation{
{
\normalsize
  \institution{School of Computer Science}
  \institution{Fudan University
  \city{Shanghai}
  \country{China}}
}
}
\email{fywang18@fudan.edu.cn}

\author{Yingxu Wang}
\authornotemark[1]
\affiliation{
{
\normalsize
  \institution{School of Computer Science}
  \institution{Fudan University
  \city{Shanghai}
  \country{China}}
}
}
\email{yingxuwang20@fudan.edu.cn}

\author{Dongsheng Li}
\affiliation{
{
\normalsize
  \institution{Microsoft Research Asia}
  \city{Shanghai}
  \country{China}
 }
}
\email{dongsli@microsoft.com}

\author{Hansu Gu}
\authornote{Corresponding author.}
{
\normalsize
\affiliation{
  \city{Seattle}
  \country{United States}}
}
\email{hansug@acm.org}

\author{Tun Lu}
\authornotemark[1]
\authornotemark[2]
\affiliation{
\normalsize
  \institution{School of Computer Science}
  \institution{Fudan University 
  \city{Shanghai} 
  \country{China}}
 }
\email{lutun@fudan.edu.cn}

\author{Peng Zhang}
\authornotemark[1]
\affiliation{
 \normalsize
  \institution{School of Computer Science}
  \institution{Fudan University
  \city{Shanghai}
  \country{China}
  }
 }
\email{zhangpeng_@fudan.edu.cn}

\author{Ning Gu}
\authornotemark[1]
\affiliation{
{
  \normalsize
  \institution{School of Computer Science}
  \institution{Fudan University
  \city{Shanghai}
  \country{China}}
 }
}
\email{ninggu@fudan.edu.cn}

\renewcommand{\shortauthors}{Wang, et al.}

\begin{abstract}
CTR prediction has been widely used in the real world. Many methods model feature interaction to improve their performance. However, most methods only learn a fixed representation for each feature without considering the varying importance of each feature under different contexts, resulting in inferior performance. Recently, several methods tried to learn vector-level weights for feature representations to address the fixed representation issue. However, they only produce linear transformations to refine the fixed feature representations, which are still not flexible enough to capture the varying importance of each feature under different contexts. In this paper, we propose a novel module named Feature Refinement Network (FRNet), which learns context-aware feature representations at bit-level for each feature in different contexts. FRNet consists of two key components: 1) Information Extraction Unit (IEU), which captures contextual information and cross-feature relationships to guide context-aware feature refinement; and 2) Complementary Selection Gate (CSGate), which adaptively integrates the original and complementary feature representations learned in IEU with bit-level weights. Notably, FRNet is orthogonal to existing CTR methods and thus can be applied in many existing methods to boost their performance. Comprehensive experiments are conducted to verify the effectiveness, efficiency, and compatibility of FRNet. 
\end{abstract}



\maketitle

\section{Introduction}
\label{sec:intro}

Click-through rate (CTR) prediction aims to estimate the probability of user clicking items, which has been widely used in Internet companies~\cite{wang2021dcn} and E-commerce platforms~\cite{zhou2018deep}. Accurate CTR prediction can deliver enormous business value and meanwhile improve users' satisfaction ~\cite{covington2016deep,zhou2018deep}, and thus has drawn increasing attention from the research community. 
Recently, many methods achieved huge success by modelling feature interactions to enrich feature representations~\cite{zhao2020dimension,zhao2021fint, liu2019feature, cheng2020adaptive, wu2020tfnet}. Following recent works~\cite{chen2021enhancing,wang2021masknet}, we categorize CTR prediction methods into two types: (1) traditional methods, such as factorization machines (FM) based methods~\cite{juan2016field,yu2019input,lu2020dual}, aim to model low-order cross-feature interactions; (2) deep learning-based methods, such as xDeepFM~\cite{lian2018xdeepfm}, AutoInt~\cite{song2019autoint}, and DCN-V2~\cite{wang2021dcn}, further enhance the accuracy of CTR prediction by capturing  high-order feature interactions.

Although existing feature interaction techniques have helped achieve better performance, they still suffer from an intrinsic issue: most of these methods only learn a fixed representation for each feature without considering the varying importance of each feature under different contexts. For example, consider the following two instances: \textit{\{female, white, computer, workday\}} and \textit{\{female, red, lipstick, workday\}}, the feature \textit{``female''} should have different representations based on its different 
influence in different instances when we make predictions for users. Such different feature representations of the same feature among different instances are called {\em context-aware feature representations} in this paper. Few CTR prediction methods~\cite{yu2019input, lu2020dual, huang2019fibinet} have attempted to learn vector-level weights for feature representations to address the fixed feature representation issue. However, it is unreasonable that these models only produce linear transformations to refine the fixed feature representations, which are still not flexible enough to capture the varying importance of each feature under different contexts. 

Self-attention mechanism has been used in CTR prediction methods~\cite{song2019autoint,li2020interpretable,lu2020dual}, which mainly learns the cross-feature relationships among all relevant feature pairs. However, self-attention uses normalized weights to capture the relative importance of features within the same instance, thus ignoring feature importance differences across multiple instances. Consider the following two instances: \textit{\{female, red, lipstick, workday\}} and \textit{\{female, red, lipstick, weekend\}}, where self-attention can only learn very similar representations for the feature ``female'' because the features ``weekend'' and ``workday'' may have very small attention scores with ``female'' compared with ``red'' and ``lipstick''. However, the behaviors/interests of ``female'' users may still significantly change from ``workday'' to ``weekend'' across the two instances. Therefore, as shown later in our case study, an ideal feature refinement module should identify the important cross-instance contextual information and learn significantly different representations under different contexts.

\begin{figure}
{\centering
\includegraphics[width=0.40\textwidth]{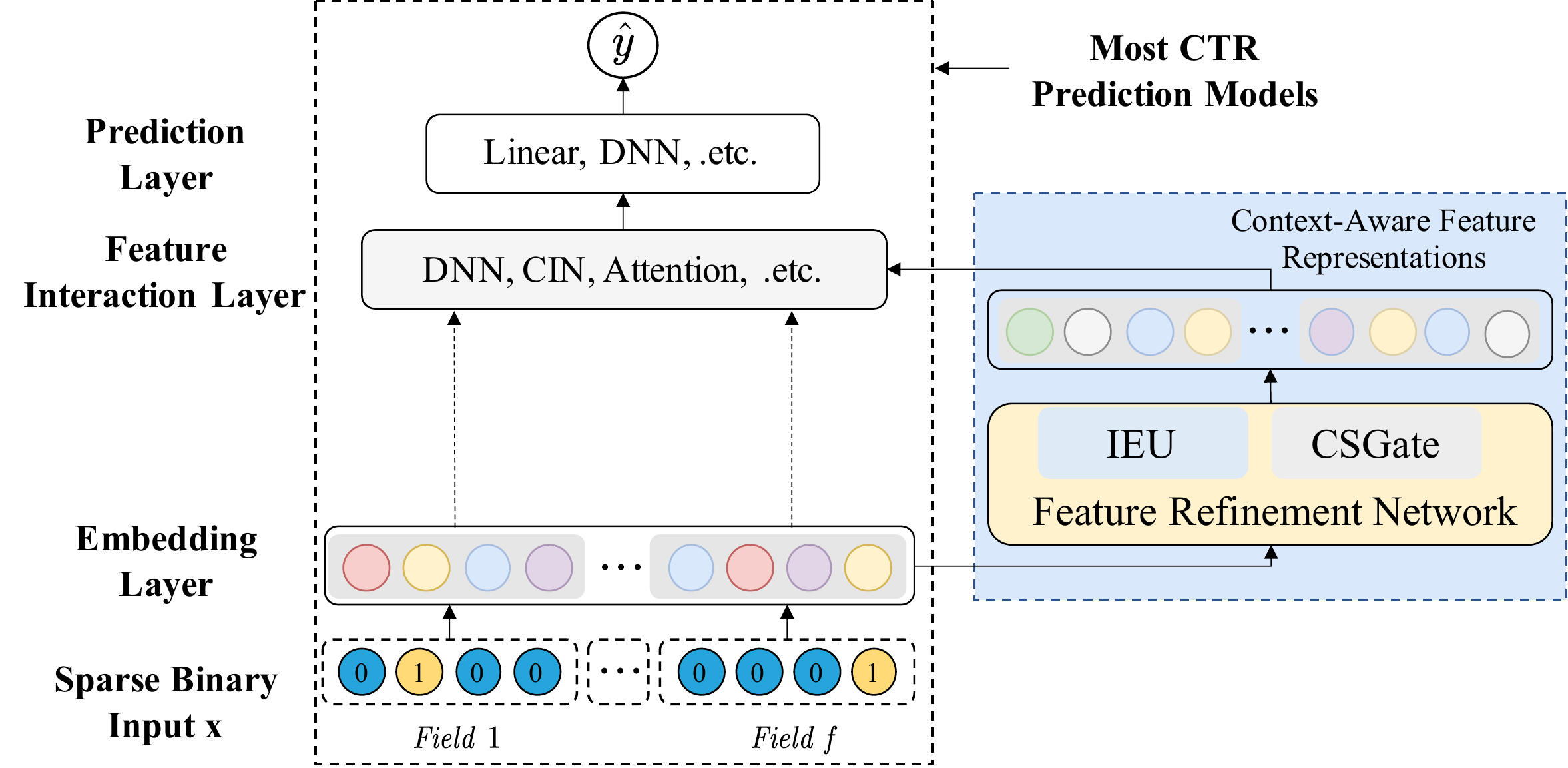} 
}\caption{General paradigm of applying the proposed Feature Refinement Network in CTR prediction methods.}
\label{Fig:paradigm}
\end{figure}

To address the above issues, we propose a novel module named \textbf{Feature Refinement Network (FRNet)} to learn context-aware feature representations. As shown in Figure~\ref{Fig:paradigm}, FRNet consists of two key components: (1) \textit{Information Extraction Unit} (IEU), which can capture contextual information and cross-feature relationships to guide context-aware feature refinement; (2) \textit{Complementary Selection Gate} (CSGate), which can adaptively integrate the original and complementary feature representations with bit-level weights to achieve context-aware feature representation learning. In IEU, we design a task-orient \textit{contextual information extractor} (CIE) to encode contextual information within each instance and employ a self-attention unit to capture the cross-feature relationships. Moreover, we design two independent IEUs in FRNet: the first IEU learns bit-level weights to select important information from the original feature representations and the second IEU generates complementary feature representations to compensate for unselected original features. In \textit{CSGate}, we design a novel gating mechanism to produce the final context-aware feature representations by integrating the original and the complementary feature representations with bit-level weights. As shown in Figure~\ref{Fig:paradigm}, FRNet is orthogonal to existing CTR prediction methods and thus can be applied in many existing methods in a plug-and-play fashion to boost their performance.

The major contributions of this paper are summarized as follows:
\begin{itemize}
    \item We propose a novel module named FRNet, which is the first work to learn context-aware feature representations by integrating the original and complementary feature representations with bit-level weights. 
    \item FRNet can be regarded as a fundamental building block to be applied in many CTR prediction methods to improve their performance. 
    \item Experimental results on four real-world datasets show that simply integrating FRNet into FM~\cite{rendle2012factorization} can outperform the state-of-the-art CTR prediction methods. Furthermore, our experiments also confirm FRNet's compatibility with many existing CTR prediction methods.
\end{itemize}

\section{Related work}

Many CTR prediction methods have achieved huge success by modeling feature interactions to enrich feature representations. Following recent works ~\cite{luo2020network,cheng2020adaptive}, we categorize CTR prediction methods into two types: traditional methods~\cite{rendle2012factorization,juan2016field,pan2018field,yu2019input,lu2020dual} and deep learning-based methods~\cite{cheng2016wide, guo2017deepfm, lian2018xdeepfm, wang2021dcn, song2019autoint,cheng2020adaptive,zhao2021fint,zhao2021non}. FM~\cite{rendle2012factorization} is one of the widely used traditional CTR prediction methods. Due to its effectiveness, many works have been proposed based on it ~\cite{juan2016field,pan2018field,yu2019input,lu2020dual}. However, these methods cannot capture high-order feature interactions. To address this issue, many deep learning-based methods were proposed to capture more complex feature interactions. Wide\&Deep (WDL)~\cite{cheng2016wide} jointly trains the wide linear unit and Multi-layer Perception (MLP) to combine the benefits of memorization and generalization. DeepFM ~\cite{guo2017deepfm} replaces the wide part of WDL with FM to alleviate manual efforts in feature engineering. Based on DeepFM, xDeepFM~\cite{lian2018xdeepfm} design a novel Compressed Interaction Network (CIN) to model high-order feature interactions explicitly. AutoInt~\cite{song2019autoint} uses stacked multi-head self-attention layers to model the feature interactions. Besides modeling feature interactions, XcrossNet~\cite{yu2021xcrossnet} and AutoDis~\cite{guo2021embedding} design various structures to learn feature embedding for numerical features. Intuitively, each feature should have different representations based on its varying roles in different instances when we make predictions. However, the above methods only learn a fixed representation for each feature without considering the varying importance of each feature under different contexts, resulting in inferior performance.

Several recent CTR prediction methods ~\cite{yu2019input, lu2020dual, huang2019fibinet} attempted to learn vector-level weights for feature representations to address the fixed feature representation issue. IFM~\cite{yu2019input} and DIFM~\cite{lu2020dual} propose Factor Estimating Network (FEN) and Dual-FEN to improve FM by learning vector-level weights for different feature representations. Similarly, FiBiNET ~\cite{huang2019fibinet} uses Squeeze-and-Excitation network (SENET)~\cite{hu2018squeeze} to extract informative features by reweighing the original features. However, only assigning vector-level weights to the same feature in different instances causes the learned representations of the same feature to have strictly linear relationships. However, it is unreasonable to only produce linear transformations to refine the fixed feature representations, because they are not flexible enough to capture the varying importance of each feature under different contexts. Recently, EGate~\cite{huang2020gatenet} applied an independent MLP for each feature to learn bit-level weights. Nevertheless, the representations of the same features are still fixed, as it only transforms the representation space.

As summarized in Table~\ref{Tab:sum}, our method is related to but fundamentally different from existing methods because we learn both bit-level weights applied in original feature embedding and complementary features to ensure that FRNet can generate more flexible nonlinear context-aware feature representations.

\begin{table}[t]
\centering
\caption{The connection and difference between FRNet and similar module. \CheckmarkBold|\XSolidBrush means totally|not met, respectively.}
\label{Tab:sum}
\scalebox{0.90}
{\small
\begin{tabular}{c|cccc} 
\hline\hline
Module (Model) & Granularity & Context-Aware & Nonlinear  \\
\hline
FEN (IFM\cite{yu2019input})                 &Vector           &\CheckmarkBold &\XSolidBrush\\
Dual-FEN (DIFM~\cite{lu2020dual})           &Vector            &\CheckmarkBold &\XSolidBrush\\
SENET (FiBiNET~\cite{huang2019fibinet})     &Vector           &\CheckmarkBold & \XSolidBrush\\
EGate (GateNet~\cite{huang2020gatenet})     & Bit      & \XSolidBrush & \XSolidBrush\\
\hline
FRNet-Vec (Ours)                             & Vector         &  \CheckmarkBold  & \CheckmarkBold  \\
FRNet (Ours)                                 & Bit            &  \CheckmarkBold  & \CheckmarkBold  \\
\hline\hline
\end{tabular}
}
\end{table}

\section{Preliminaries}
\label{sec:pre}
CTR prediction is a binary classification task on sparse multi-field categorical data~\cite{pan2021click, huang2019fibinet, yu2021xcrossnet}. Suppose there are $f$ different fields and $n$ features, each field may contain multiple features but each feature only belongs to one field. Each instance for CTR prediction can be represented by $\left\{ \mathbf{x}_i, y_i \right\}$, where $\mathbf{x}_i$ is a sparse high-dimensional vector represented by one-hot encoding and $y_i\in{\{0,1\}}$ (click or not) is the true label, e.g.,
\begin{align}
\mathbf{x}_i= \underset{Item=Computer\,\,}{\underbrace{\left( 0,...,1,0 \right) }}\underset{Color=White\,\,}{\underbrace{\left( 1,...,0,0 \right) }}\mathbf{...}\underset{Gender=Female}{\underbrace{\left( 1,0 \right). }}
\end{align}
CTR prediction models aim to approximate the probability $P(y_i|\mathbf{x}_i)$ for each instance. According to~\cite{wei2021autoias,wang2021masknet}, most recent CTR prediction methods follow the design paradigm below (as shown in Figure~\ref{Fig:paradigm}):

\textbf{\textit{Embedding layer.}} It transforms the sparse high-dimensional features $\mathbf{x}_i$ into a dense low-dimensional embedding matrix $\mathbf{E}=[\mathbf{v}_1;\mathbf{v}_2;...;\mathbf{v}_f] \in \mathbb{R}^{f\times d}$, where $d$ is the dimension size of each field. Each feature has a fixed-length representation $\mathbf{v}_i$. 

\textbf{\textit{Feature interaction layer.}} In CTR prediction methods, the most critical design is the feature interaction layer, which uses various types of interaction operations to capture arbitrary-order feature interactions, such as MLP~\cite{guo2017deepfm,cheng2016wide}, Cross Network~\cite{wang2021dcn,wang2017deep} and transformer layer~\cite{li2020interpretable,song2019autoint}, etc. The output of feature interaction layer is a compact representation $\mathbf{q}_i$ based on embedding matrix $\mathbf{E}$. 

\textbf{\textit{Prediction layer.}} Finally, a prediction layer (usually a linear regression or MLP module) produces the final prediction probability $\sigma(\hat{y_i}) \in (0, 1)$ based on the representations $\mathbf{q}_i$, where $\sigma(x)=1 /(1+\exp (-x))$ is the sigmoid function. And, a common loss function for CTR prediction tasks is the cross entropy loss as follows:
\begin{align}
\textstyle
loss =-\frac{1}{N}\sum_{i=1}^N{y_i}  \log\left( \sigma \left(\hat{y}_i \right)\right) +\left( 1-y_i \right) \log \left( 1-\sigma \left(\hat{y}_i \right)\right),
\end{align}
where $N$ is total number of training instances.

\begin{figure*}[t]
{\centering 
\includegraphics[width=0.75\textwidth]{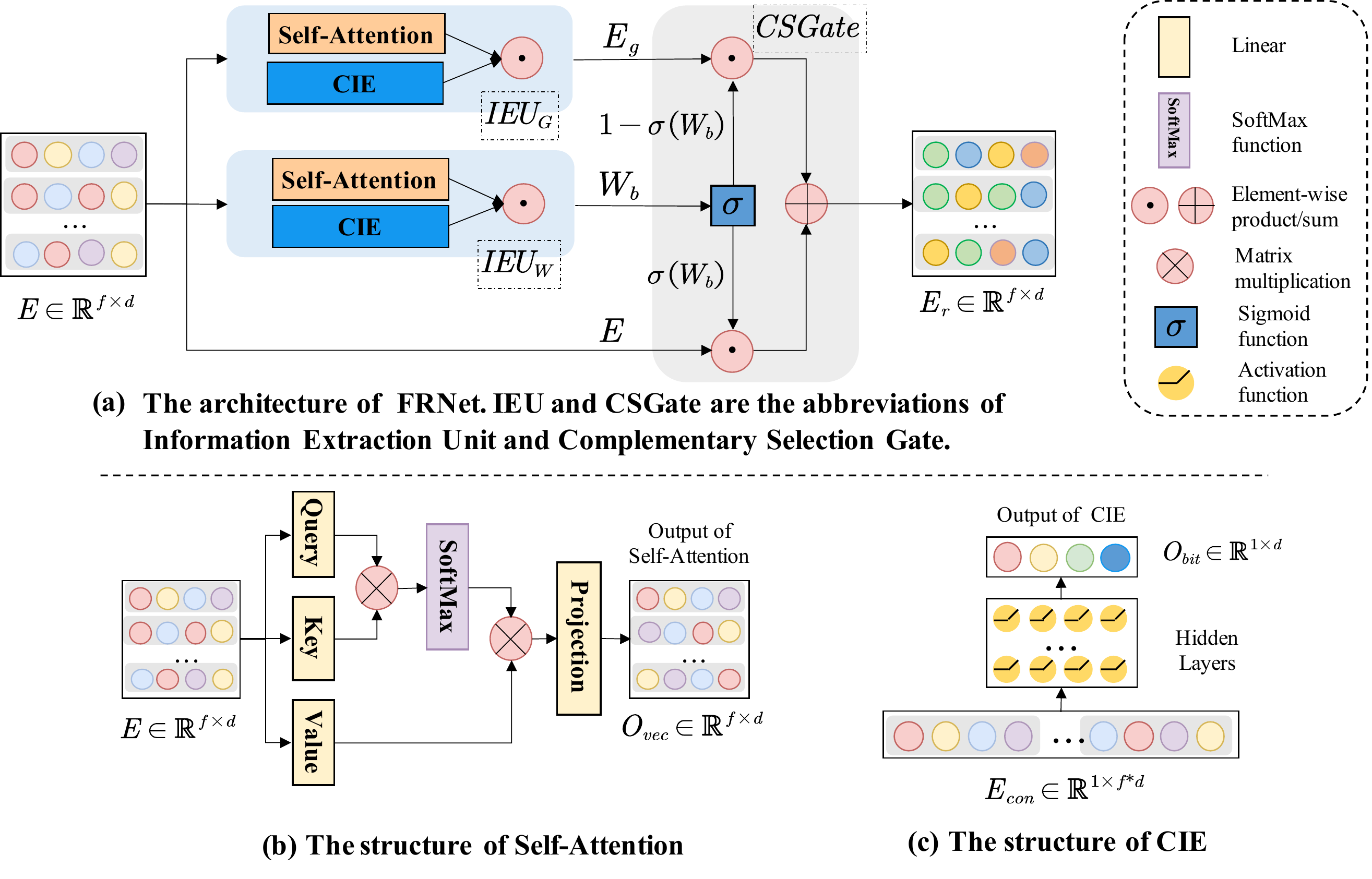}
}\caption{Architectures of Feature Refinement Network (FRNet), Self-Attention and Contextual Information Extractor (CIE).}
\label{Fig:frnet01} 
\end{figure*}

\section{Feature Refinement Network}\label{model:frnet}
In this section, we introduce the details of FRNet. As depicted in Figure~\ref{Fig:frnet01} (a), FRNet contains two key components:
\begin{itemize}
\item \textit{Information Extraction Unit} (IEU), which can capture contextual information and cross-feature relationships to guide context-aware feature refinement.
\item \textit{Complementary Selection Gate} (CSGate),  which can adaptively integrate the original and complementary feature representations with bit-level weights to achieve context-aware feature representation learning.
\end{itemize}

\subsection{Information Extraction Unit (IEU)} 

IEU consists of three essential components: 1) the Self-Attention unit, which is deployed to capture explicit cross-feature relationships among co-occurring features; 2) Contextual Information Extractor (CIE), which aims to encode the contextual information under different contexts; and 3) Integration unit, which integrates the information from the Self-Attention unit and CIE. In addition, we use two IEUs for two purposes: $IEU_W$ learns bit-level weights, and $IEU_G$ produces complementary feature representations.

\subsubsection{Self-Attention unit} 
We adopt self-attention~\cite{vaswani2017attention} to identify the most relevant features to each specific feature in instances. For instance, in \textit{\{female, red, lipstick, workday\}}, the most relevant features to ``\textit{female}'' are ``\textit{red}'' and ``\textit{lipstick}''. The self-attention module first calculates importance among all feature pairs and generates new representations by computing the weighted sum of relevant features. To achieve higher efficiency, we simplify the structure of self-attention as depicted in Figure~\ref{Fig:frnet01} (b). More detailed, we first map the input matrix $\mathbf{E}$ into three different matrices: 
\begin{align}
    \mathbf{Q},\mathbf{K},\mathbf{V} = \mathbf{EW^Q}, \mathbf{EW^K}, \mathbf{EW^V},
\end{align}
where $\mathbf{W^Q}$, $\mathbf{W^K}$, $\mathbf{W^V} \in \mathbb{R}^{d \times d_k}$ are transformation matrices, and $d_k$ is the attention size. Then, we obtain the attention matrix on Value ($\mathbf{V}$) by applying the dot product of  Query ($\mathbf{Q}$) and Key ($\mathbf{K}$) with a Softmax function as follows:
\begin{align}
    Attention(\mathbf{Q},\mathbf{K},\mathbf{V}) = SoftMax(\mathbf{QK^T})\mathbf{V} \in \mathbb{R}^{f \times d_k}.
\end{align}
Finally, we transform the dimension of output matrix to be the same as the input by a projection matrix  $\mathbf{W^P} \in \mathbb{R}^{d_k \times d}$ . The output ($\mathbf{O_{vec}}$) of the self-attention module can be summarized as follows:
\begin{align}
    \mathbf{O_{vec}} = Attention(\mathbf{Q,K,V})\mathbf{W^{P}} \in \mathbb{R}^{f \times d}. \label{equ:att}
\end{align}
The self-attention mechanism can achieve partially context-aware feature representation learning by capturing the cross-feature relationships among all feature pairs to refine the feature representation under different contexts. However, self-attention only utilizes partial contextual information represented by pair-wise feature interactions and thus fails to utilize complete contextual information to guide feature refinement. In other words, self-attention yields similar feature representations for the same features in different instances, as shown in our studies (Section~\ref{sec:attention_analysis}).

\subsubsection{Contextual Information Extractor}
\label{sec:cie}

The contextual information in each instance is implicitly contained in all features. Hence, we need to ensure that all features contribute to the contextual information in each instance. Since the contextual information is usually not very complicated, MLP is a simple yet effective choice to extract contextual information as shown in the experiments (Section~\ref{sec:hp}). In detail, we first concatenate the original feature representations into $\mathbf{E_{con}}$ as the input. 

Then, each layer of the MLP is obtained as follows:
\begin{align}
    \mathbf{h}_{l+1} = PReLU(\mathbf{W}_{l}\mathbf{h}_{l}+b_{l}),
\end{align}
where $\mathbf{h}_{l} \in \mathbb{R}^{n_l}$, $\mathbf{h}_{l+1} \in \mathbb{R}^{n_{l+1}}$ are the $l$-th and ${(l+1)}$-th hidden layer, and $\mathbf{h}_0=\mathbf{E_{con}} \in \mathbb{R}^{1 \times (f*d)}$. $\mathbf{W}_{l} \in \mathbb{R}^{n_{l+1}\times n_{l}}$, $b_{l}$ are the learnable parameters for the $l$-th deep layer. PReLU(·) is the PReLU~\cite{he2015delving} function. In the last hidden layer, we project the dimension of contextual information vector to $d$ (the dimension of embedding size), and compute the contextual information vector $\mathbf{O_{bit}}$ as follows:
\begin{align}
    \mathbf{O_{bit}} = PReLU(\mathbf{W}_{L}\mathbf{h}_{L}+b_{L})  \in \mathbb{R}^{1 \times d} , \label{equ:prelu2}
\end{align}
where $\mathbf{W}_{L} \in \mathbb{R}^{ d \times n_{L-1}}$, $b_{L}$ are the parameters of the last layer. Since $\mathbf{O}_{bit}$ compresses all information from $\mathbf{E_{con}}$, it can represent the contextual information within the specific instance. Intuitively, contextual information $\mathbf{O_{bit}}$ is unique for each instance, as different instances contain different features. 

\subsubsection{Integration unit} 
\begin{figure}
{\centering 
\includegraphics[width=0.45\textwidth]{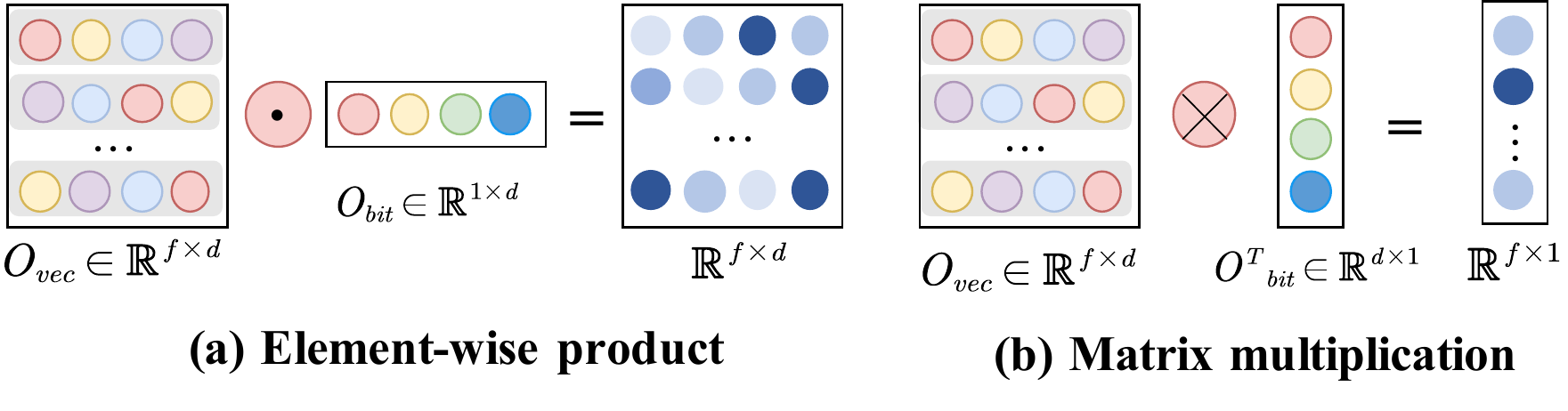} 
}\caption{The operations to integrate the outputs of self-attention and CIE units. } 
\label{Fig:product} 
\end{figure}

After obtaining the contextual information $\mathbf{O_{bit}}$, we directly use $\mathbf{O_{bit}}$ to weigh the feature representation  $\mathbf{O_{vec}}$. As illustrated in Figure~\ref{Fig:product} (a), it is calculated as follows:
\begin{align}
    \mathbf{O_{IEU}} =  \mathbf{O_{vec}} \odot \mathbf{O_{bit}}  \in \mathbb{R}^{f\times d}.
    \label{eqn:O_IEU}
\end{align}
$\odot$ is the element-wise product. $\mathbf{O_{vec}}$ is the feature representation from self-attention module which captures cross-feature relationships, and $\mathbf{O_{bit}}$ enables each feature representation to be aware of the contextual information. Equation~\ref{eqn:O_IEU} ensures each feature can have significantly different representations in different instances.

As shown in Figure~\ref{Fig:frnet01} (a), we deploy two independent IEUs: (1) $IEU_W$ learns bit-level weights; (2) $IEU_G$ produces complementary features. Specifically, their outputs are represented as follows:
\begin{gather}
    \mathbf{E_{g}} = \mathbf{IEU_G(E)} \in \mathbb{R}^{f\times d},~~
    \mathbf{W_{b}} = \mathbf{IEU_W(E)} \in \mathbb{R}^{f\times d}.
    \label{eqn:IEU_func}
\end{gather}
We will present more details of Equation~\ref{eqn:IEU_func} in the next subsection.

\subsection{Complementary Selection Gate (CSGate)}
\label{sec:csgate}
In CSGate, we design a novel gate mechanism to control information flow and select important information from the original and complementary features with bit-level weights. As shown in Figure \ref{Fig:frnet01} (a), CSGate has three different inputs from three channels: 1) complementary feature representations $\mathbf{E_g}$; 2) weight matrix $\mathbf{W_{b}}$\footnote{Although $\sigma (\mathbf{W_{b}})$ is the real weight matrix, we also call $\mathbf{W_{b}}$ as weight matrix for the sake of convenience. Similarly, we also call $\mathbf{E_{g}} $ as complementary features.}; and 3) original feature representations $\mathbf{E}$. The output of CSGate is the context-aware feature representation:
\begin{align}
\mathbf{E}_{\mathbf{r}}=\underset{Selected\,\,features\,\,}{\underbrace{\mathbf{E}\odot \sigma \left( \mathbf{W_{b}} \right) }}+\underset{Complementary\ features\,\,}{\underbrace{\mathbf{E}_{\mathbf{g}}\odot \left( 1-\sigma \left( \mathbf{W_{b}} \right)\right) }},
\label{equ:gate}
\end{align}
where $\sigma(\cdot)$ is the sigmoid function. $\mathbf{E_{r}} \in \mathbb{R}^{f\times d}$ has the same dimensions as $\mathbf{E}$. Specifically, Equation \eqref{equ:gate} contains two parts of features: \textit{selected features} and \textit{complementary features}. Those two parts are connected by the \textit{selection gate} $\sigma(\mathbf{W_{b}})$.

\textbf{\textit{Selected Features}} are the selected important information from original feature representations at bit-level. Specifically, each element in $\sigma(\mathbf{W_{b}})$ measures the importance of the corresponding element in the original feature representation $\mathbf{E}$, where the probability of selecting the specific element is between 0 and 1. Thus, we can learn nonlinear context-aware feature representations. Compared with previous works~\cite{lu2020dual,yu2019input,huang2019fibinet}, the learned weight matrix $\sigma(\mathbf{W_{b}})$ have two advantages: 1) it contains cross-feature relationships and contextual information simultaneously, which enables to learn context-aware representations for the same feature in different instances; and 2) introducing the bit-level weights to the original feature representations can achieve more flexible and fine-grained feature refinement than previous linear transformations.

\textbf{\textit{Complementary Features}} are the complementary information that aims to further enhance the expressive capacity of context-aware feature representations. Existing methods~\cite{yu2019input, huang2019fibinet,lu2020dual} only assign weights to original features without considering the unselected information. However, we believe the unselected features may still help CTR prediction in a different way. Hence, we propose to leverage $\mathbf{E_g}$ with its weight $1-\sigma (\mathbf{W_{b}})$ as the other part of the final context-ware feature representations. In particular, the gate $\sigma (\mathbf{W_{b}})$ achieves adaptive balance between the selected features and complementary features in bit level.

In summary, FRNet generates context-aware feature representation by three steps:
1) generating complementary feature representations using $IEU_G$;
2) calculating bit-level weight matrix by $IEU_W$; and 
3) leveraging the CSGate to generate context-aware feature interactions by integrating original features representations and complementary feature representations by bit-level weights.

\section{Experiments}

\subsection{Experimental Setup}

\subsubsection{Datasets}
We conduct experiments on four popular datasets:

\textbf{Criteo}\footnote{\url{https://www.kaggle.com/c/criteo-display-ad-challenge}} is the most well-known industrial benchmark dataset for CTR prediction, which includes 26 anonymous categorical fields and 13 numerical fields. We discretize numerical features and transform them into categorical features by log transformation\footnote{\url{https://www.csie.ntu.edu.tw/~r01922136/kaggle-2014-criteo.pdf}}. And following ~\cite{yang2020operation}, we use the last 5 million records for testing. Meanwhile, we remove the features appeared less than 10 times and treat them as a dummy feature ``$\langle unknown \rangle$''.

\textbf{Malware}\footnote{\url{https://www.kaggle.com/c/microsoft-malware-prediction}} is published in the Microsoft Malware prediction, which contains 81 different fields. This task can be transformed as a binary classification problem like a CTR prediction task~\cite{wang2021masknet}. 

\textbf{Frappe}\footnote{\url{https://www.baltrunas.info/context-aware/frappe}} contains app usage logs from users under different contexts (e.g., daytime, location). The target value indicates whether the user has used the app under the context~\cite{xiao2017attentional}.

\textbf{MovieLens}\footnote{\url{https://grouplens.org/datasets/movielens/}} contains user tagging records on movies. Each instance contains three fields: user ID, movie ID, tag. The targeted value denotes whether a user has assigned a tag to a movie~\cite{xiao2017attentional}. 

The statistics of these four datasets are summarized in Table \ref{Tab.dataset}. 

\subsubsection{Evaluation Metrics}
To evaluate the performance of CTR prediction methods, we adopt \textbf{AUC} (Area under the ROC curve)  and \textbf{Logloss} (binary cross-entropy loss) as the evaluation metrics~\cite{chen2021enhancing, wang2021masknet}. Note that slightly higher AUC or lower Logloss, e.g., at \textbf{0.001} level, can be regarded as significant improvement in CTR prediction tasks~\cite{chen2021enhancing,luo2020network, wang2021dcn,cheng2016wide, huang2019fibinet,lian2018xdeepfm,li2020field}. 


\subsubsection{Compared Models}
We apply FRNet into FM~\cite{rendle2012factorization}, which is called $FM_{FRNet}$. We compare $FM_{FRNet}$ with three types of methods: 
1) FM-based methods, which capture second- or higher-order feature interactions, including \textbf{FM}~\cite{rendle2012factorization}, \textbf{IFM}~\cite{yu2019input}, \textbf{DIFM}~\cite{lu2020dual}; 
2) deep learning-based methods, which model high-order feature interactions, including \textbf{NFM}~\cite{he2017neural},  \textbf{IPNN}~\cite{qu2018product}, \textbf{OPNN}~\cite{qu2018product},  \textbf{CIN}~\cite{lian2018xdeepfm}, \textbf{FINT}~\cite{zhao2021fint}; and 
3) ensemble methods, which adopt multi-tower feature interaction structures to integrate different types of methods, including \textbf{WDL}~\cite{cheng2016wide}, \textbf{DCN}~\cite{wang2017deep}, \textbf{DeepFM}~\cite{guo2017deepfm}, \textbf{xDeepFM}~\cite{lian2018xdeepfm}, \textbf{FiBiNET}~\cite{huang2019fibinet}, \textbf{AutoInt+}~\cite{song2019autoint},   \textbf{AFN+}~\cite{cheng2020adaptive}, \textbf{NON}~\cite{luo2020network}, \textbf{TFNET}~\cite{wu2020tfnet}, \textbf{FED}~\cite{zhao2020dimension}, and \textbf{DCN-V2}~\cite{wang2021dcn}. 
We do not present the results of   classical methods, including \textbf{LR}~\cite{richardson2007predicting}, \textbf{GBDT}~\cite{he2014practical}, \textbf{CCPM}~\cite{gehring2017convolutional}, \textbf{FFM}~\cite{juan2016field}, \textbf{AFM}~\cite{xiao2017attentional}, \textbf{FwFM}~\cite{pan2018field}, \textbf{CrossNet}~\cite{wang2017deep}, \textbf{FNN} ~\cite{zhang2016deep}, because more recent methods (e.g., AFN+~\cite{cheng2020adaptive}, FiBiNET~\cite{huang2019fibinet}, DCN-V2~\cite{wang2021dcn}) have outperformed these methods in their experiments.

To demonstrate the effectiveness of the bit-level weights in FRNet, we design a variant of FRNet named \textbf{FRNet-Vec}, where {FRNet-Vec} only learns the vector-level weights in IEU and keep the other parts the same as FRNet. As shown in Figure~\ref{Fig:product} (b), the weight matrix $\mathbf{W}_{v}$ in FRNet-Vec is calculated by:
\begin{align}
\mathbf{W}_{v} =  \mathbf{O_{vec}} \otimes \mathbf{O^T_{bit}} \in \mathbb{R}^{f\times 1}.
\end{align}
Each element in $\sigma(\mathbf{W}_{v})$  measures the importance of each feature representations in original embedding $\mathbf{E}\in \mathbb{R}^{f\times d}$.

\begin{table}[t] 
\centering
\caption{Statistics of four datasets used in this paper.}
\label{Tab.dataset}
\scalebox{0.75}{
\begin{tabular}{c|cccccc} 
\hline\hline
Datasets & Positive  & \#Training & \#Validation & \#Testing & \#Fields & \#Features   \\ 
\hline
Criteo  & 26\%  & 35,840,617      & 5,000,000  & 5,000,000  &39  & 1,086,810  \\
Malware  & 50\%   &7,137,187     &892,148        &892,148   &81   & 976,208  \\
Frappe    & 33\%  & 202,027       & 57,722     & 28,860     & 10  & 5,382      \\
MovieLens & 33\%  & 1,404,801     & 401,372    & 200,686     & 3   & 90,445     \\
\hline\hline
\end{tabular}
}
\end{table}
\subsubsection{Implementation Details.}
We implement our method with Pytorch\footnote{The code is available here: \url{https://github.com/frnetnetwork/frnet}}. All models are learned by optimizing the Cross-Entropy loss with Adam~\cite{kingma2014adam} optimizer. We implement the Reduce-LR-On-Plateau scheduler during the training process to reduce the learning rate by a factor of 10, when the given metric stops improving in four consecutive epochs. The default learning rate is 0.001. We use early stop to avoid overfitting when the AUC on the validation set stops improving. The mini-batch size is set to 4096. The embedding size is 10 for Criteo and Malware and 20 for Frappe and MovieLens, respectively. Following previous works~\cite{huang2019fibinet,cheng2020adaptive, guo2017deepfm, song2019autoint}, we employ the same neural structure (i.e., 3 layers, 400-400-400) for the models that involve MLP for a fair comparison. All activation functions are ReLU unless otherwise specified, and the dropout rate is set to 0.5. In FRNet, the dimension of MLP in the CIE is set to 128. For other methods, we take the optimal settings from the original papers.

To ensure fair comparison, we run all experiments five times by changing random seeds and report the averaged results. We observe that all the standard deviations of our method are in the order of \textbf{1e-4}, indicating that our results are very stable. We further perform two-tailed t-test to verify the statistical significance in comparisons between our method and the best baseline methods.

\begin{table*}[t]
\centering
\caption{Overall accuracy comparison in the four datasets. $\Delta_{AUC}$ and $\Delta_{Logloss}$ are calculated to indicate averaged performance boost compared with the best baseline (DCN-V2) over the four datasets. Two tailed t-test: $\star: p < 10^{-2}$,  $\star \star : p < 10^{-4}$. 
}
\label{tab:all}
\scalebox{0.85}{
\begin{tabular}{cc|cc|cc|cc|cc|cc} 
\hline \hline

\multirow{2}{*}{\begin{tabular}[c]{@{}c@{}}Model\\Class\end{tabular}}
& Datasets& \multicolumn{2}{c|}{Criteo} & \multicolumn{2}{c|}{Malware} &\multicolumn{2}{c|}{Frappe}& \multicolumn{2}{c}{MovieLens}&
\multirow{2}{*}{\begin{tabular}[c]{@{}c@{}}$\Delta_{AUC} $\\ $\uparrow$ \end{tabular}} &
\multirow{2}{*}{\begin{tabular}[c]{@{}c@{}}$\Delta_{Logloss} $\\$\downarrow$\end{tabular}} \\

\cline{2-10}
& Model& AUC& Logloss& AUC& Logloss& AUC& Logloss& AUC& \multicolumn{1}{c}{Logloss} & \multicolumn{1}{l}{}& \multicolumn{1}{l}{}\\ 

\hline
\multirow{3}{*}{\begin{tabular}[c]{@{}c@{}}Second-Order\end{tabular}} 
& FM& 0.8028& 0.4514& 0.7309& 0.6052& 0.9708& 0.1934& 0.9391& 0.2856& -1.13\%& +0.0167\\
& IFM    & 0.8066& 0.4470& 0.7389& 0.5969& 0.9765& 0.1896& 0.9471& 0.2853& -0.39\%& +0.0125\\
& DIFM   & 0.8085& 0.4457& 0.7397& 0.5954& 0.9788& 0.1860& 0.9490& 0.2459& -0.19\%& +0.0011\\ 
\hline
\multirow{5}{*}{\begin{tabular}[c]{@{}c@{}}High-Order\end{tabular}} 
& NFM    & 0.8057& 0.4483& 0.7352& 0.5988& 0.9746& 0.1915& 0.9437& 0.2945& -0.68\%& +0.0161\\
& IPNN  & 0.8088& 0.4454& 0.7404& 0.5945& 0.9791& 0.1759& 0.9490& 0.2785& -0.15\%& +0.0064\\
& OPNN  & 0.8096& 0.4446& 0.7408& 0.5840& 0.9795& 0.1805& 0.9497& 0.2704& -0.08\%& +0.0027\\
& CIN   & 0.8082& 0.4459& 0.7395& 0.5967& 0.9776& 0.2010& 0.9483& 0.2808& -0.26\% & +0.0139\\
& FINT  & 0.8090    & 0.4452& 0.7402& 0.5953& 0.9791    & 0.1921    & 0.9498 & 0.2674  & -0.13\%& +0.0078\\ 
\hline
\multirow{11}{*}{\begin{tabular}[c]{@{}c@{}}Ensemble\end{tabular}}    
& WDL   & 0.8068& 0.4474& 0.7392& 0.5982& 0.9776& 0.1895& 0.9403& 0.3045& -0.52\% & +0.0177\\
& DCN   & 0.8091& 0.4452& 0.7403&0.5944 & 0.9789& 0.1814& 0.9458 & 0.2685 & -0.23\%& +0.0052\\
& FiBiNET& 0.8093& 0.4450& 0.7405& 0.5942& 0.9787& 0.1867& 0.9471& 0.2630& -0.19\%& +0.0050\\
& DeepFM & 0.8084& 0.4458& 0.7402& 0.5944& 0.9789& 0.1770& 0.9465& 0.3079& -0.24\%& +0.0141\\
& xDeepFM& 0.8086& 0.4456& 0.7405& 0.5940& 0.9792& 0.1889& 0.9480& 0.2889& -0.18\%& +0.0122\\
& AutoInt+ & 0.8088& 0.4456& 0.7406& 0.5939& 0.9786& 0.1890& 0.9501& 0.2813& -0.13\%& +0.0103\\
& AFN+  & 0.8095& 0.4447& 0.7404& 0.5945& 0.9791& 0.1824& 0.9509& 0.2583& -0.08\%& +0.0028\\
& NON   & 0.8096& 0.4446& 0.7390& 0.5956& 0.9792& 0.1813& 0.9505& 0.2625& -0.13\%& +0.0038\\
& TFNet & 0.8092& 0.4449& 0.7397& 0.5948& 0.9787& 0.1942& 0.9493& 0.2714 & -0.16\%& +0.0091\\
& FED   & 0.8087& 0.4458& 0.7406& 0.5942& 0.9797& 0.1802& 0.9510 & 0.2576  & -0.08\%& +0.0022\\
& DCN-V2& 0.8098    & 0.4443    & 0.7411    & 0.5935    & 0.9802    & 0.1783    & 0.9516 & 0.2527  & - & - \\
\hline
\multirow{2}{*}{\begin{tabular}[c]{@{}c@{}}Our\\Models\end{tabular}} 
& $FM_{FRNet-Vec}$ & $0.8115^{\star\star}$& $0.4428^{\star\star}$& $0.7438^{\star\star}$& $0.5914^{\star\star}$& $0.9816^{\star\star}$& $0.1653^{\star}$& $0.9635^{\star\star}$& $0.2365^{\star\star}$& 0.49\%& -0.0082\\
& $FM_{FRNet}$& $\textbf{0.8120}^{\star\star}$ & $\textbf{0.4424}^{\star\star}$ & $\textbf{0.7445}^{\star\star}$ & $\textbf{0.5909}^{\star\star}$ & $\textbf{0.9830}^{\star\star}$ & $\textbf{0.1607}^{\star\star}$ & $\textbf{0.9679}^{\star\star}$ & $\textbf{0.2278}^{\star}$   & \textbf{0.68}\% & \textbf{-0.0118}\\
\hline
\hline
\end{tabular}
}
\end{table*}
\begin{figure*}[t]

{\centering 
\includegraphics[width=0.90\textwidth]{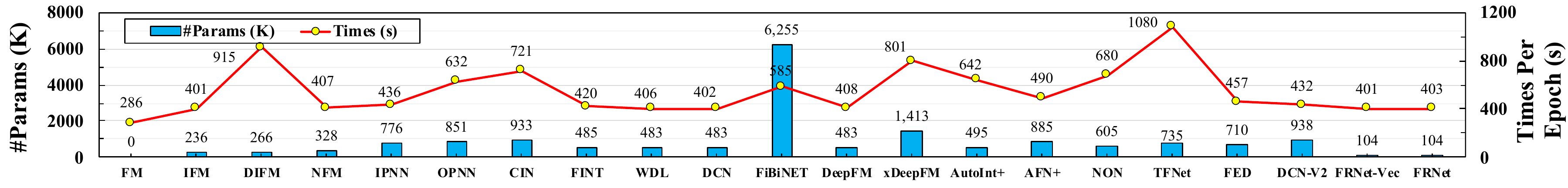}
}\caption{Efficiency comparisons of different algorithms in terms of \textit{\textbf{Model Size}} and \textit{\textbf{Run Time}} on the Criteo dataset. We only consider the parameters above the embedding layer (including LR part). We conduct experiments with one TITAN V GPU.}
\label{fig:time} 
\end{figure*}

\subsection{Overall Comparison}
\label{sec:comparison}
\subsubsection{Effectiveness Comparison.}
Table~\ref{tab:all} summarizes the effectiveness of FRNet and all compared methods on the four datasets. 
Although FM has the worst performance, $FM_{FRNet}$ and $FM_{FRNet-Vec}$ statistically significantly outperform all compared methods. Specifically, $FM_{FRNet}$ outperforms FM by 1.15\%, 1.86\%, 1.26\% and 3.07\% in terms of AUC (1.99\%, 2.36\%, 16.91\% and 20.23\% in terms of Logloss) on four datasets, respectively, which demonstrates that learning context-aware feature representations is effective in CTR prediction. Meanwhile, the averaged performance boost ($\Delta_{AUC}$ and $\Delta_{Logloss}$) indicate the most strong generalization ability of $FM_{FRNet}$ and $FM_{FRNet-Vec}$ on four datasets. In addition, $FM_{FRNet}$ achieves better performance than $FM_{FRNet-Vec}$, which confirms that refining feature representations at bit-level is more effective. Most importantly, Table \ref{tab:all} indicates that learning context-aware feature representations by FRNet is more effective than other feature interaction techniques, e.g., the ones in xDeepFM, NON, and DCN-V2.

\subsubsection{Efficiency Comparison.}
We compare the \textit{model size} and \textit{run time} of different methods in Figure~\ref{fig:time}. Generally, FM-based methods have fewer parameters than high-order or ensemble methods. Specifically, $FM_{FRNet}$ only increases 104K learning parameters over FM. As a comparison, DIFM and xDeepFM increase 266K and 483K learning parameters over FM, respectively. Meanwhile, they are relatively time-consuming, as they consist of complicated structures, e.g., Dual-FEN and CIN. We also observe from Figure~\ref{fig:time} that $FM_{FRNet}$ is comparable to IFM and DCN, and has fewer model parameters and is more efficient than all other baseline methods. Notably, compared with the best-performing baseline  DCN-V2, $FM_{FRNet}$ has fewer model parameters, faster training speed and better performance.

\begin{table*}[t]
\centering
\caption{Compatibility comparison between FRNet and other four modules over SOTA CTR prediction methods.}
\label{Tab:module}
\scalebox{0.90}{
\begin{tabular}{cc|cc|cc|cc|cc|cc|cc} 
\hline\hline
\multirow{2}{*}{Datasets}& Modules & \multicolumn{2}{c|}{BASE} & \multicolumn{2}{c|}{SENET (FiBiNET)} & \multicolumn{2}{c|}{EGate (GateNet)} & \multicolumn{2}{c|}{FEN (IFM)} & \multicolumn{2}{c|}{Dual-FEN (DIFM)} & \multicolumn{2}{c}{FRNet (Ours)} \\ 
\cline{2-14}

& Models& AUC& Logloss& AUC& Logloss& AUC    & Logloss& AUC    & Logloss& AUC    & Logloss& AUC& Logloss\\ 
\hline
\multirow{8}{*}{\rotcell{Criteo}}
& FM& 0.8028 & 0.4514& 0.8073& 0.4467& 0.8058 & 0.4482& 0.8066 & 0.4470& 0.8085 & 0.4457& \textbf{0.8120}  & \textbf{0.4424}  \\

 & AFM& 0.7999 & 0.4535& 0.8048& 0.4486& 0.7925 & 0.4601& 0.7951 & 0.4576& 0.7924   & 0.4600& \textbf{0.8116}& \textbf{0.4427} \\
& NFM& 0.8057 & 0.4483& 0.8063& 0.4476& 0.8060 & 0.4479& 0.8063 & 0.4474& 0.8080 & 0.4461& \textbf{0.8120}  & \textbf{0.4425}   \\

& DeepFM& 0.8084 & 0.4458& 0.8089& 0.4453 & 0.8085 & 0.4457& 0.8085 & 0.4459    & 0.8094 & 0.4448& \textbf{0.8118}  & \textbf{0.4426}   \\
& xDeepFM   & 0.8086 & 0.4456& 0.8093& 0.4451 & 0.8100 & 0.4442& 0.8087 & 0.4455    & 0.8101 & 0.4443& \textbf{0.8110}  & \textbf{0.4434}   \\
& IPNN & 0.8088 & 0.4454& 0.8100& 0.4442 & 0.8104 & 0.4438& 0.8102 & 0.4441    & 0.8094 & 0.4450& \textbf{0.8115} & \textbf{0.4428}   \\
& FiBiNET   & 0.8093 & 0.4450& 0.8093& 0.4450 & 0.8102 & 0.4440& 0.8102 & 0.4439    & 0.8104 & 0.4436& \textbf{0.8119}  & \textbf{0.4425}   \\
\cline{2-14}
& \textit{Avg. Imp}   & - & - &0.22\%&	0.40\%&	0.00\%&	0.04\%	&0.04\%	&0.12\%	&0.08\%	&0.18\%	& \textbf{0.68\%}	&\textbf{1.15\%}\\

\hline
\multirow{8}{*}{\rotcell{Frappe}}  
& FM& 0.9708 & 0.1934& 0.9764& 0.1863 & 0.9515 & 0.3134 & 0.9765 & 0.1896& 0.9788 & 0.1860& \textbf{0.9830}&\textbf{0.1607}   \\
& AFM  & 0.9606 & 0.2483& 0.9620&0.2453&0.9477&0.2733 &0.9487&0.2704 &0.9698&0.2417 & \textbf{0.9803} & \textbf{0.1831}\\
& NFM  & 0.9746 & 0.1915& 0.9787& 0.1794 & 0.9754 & 0.1860& 0.9774 & 0.1778    & 0.9785 & 0.1758& \textbf{0.9822}  & \textbf{0.1620}   \\

& DeepFM    & 0.9789 & 0.1770& 0.9813& 0.1642 & 0.9808 & 0.1682& 0.9817 & 0.1625    & 0.9796 & 0.1727& \textbf{0.9836}  & \textbf{0.1594}   \\
& xDeepFM   & 0.9792 & 0.1889& 0.9817& 0.1629 & 0.9805 & 0.1694& 0.9814 & 0.1679    & 0.9807 & 0.1715& \textbf{0.9824}  & \textbf{0.1653}   \\
& IPNN & 0.9791 & 0.1759& 0.9812& 0.1639 & 0.9805 & 0.1667& 0.9815 & 0.1634    & 0.9809 & 0.1650& \textbf{0.9828}  &\textbf{0.1597}   \\

& FiBiNET   & 0.9787 & 0.1867& 0.9787 & 0.1867   & 0.9803 & 0.1674  & 0.9798 & 0.1736    & 0.9805 & 0.1648& \textbf{0.9821}  & \textbf{0.1635}   \\
\cline{2-14}
& \textit{Avg. Imp}  & - & - & 0.27\% &8.64\% &	-0.37\% &	-2.31\% &	0.07\% &	7.84\% &	0.40\% &	9.32\% &	\textbf{0.80\%} &	\textbf{17.43\%}    \\
\hline\hline
\end{tabular}
}
\end{table*}

\subsection{Compatibility Analysis}

\label{sec:compatibility}
To confirm the compatibility of FRNet, we apply FRNet in seven CTR prediction methods. Meanwhile, we compare FRNet with additional four modules proposed by recent works which assign different weights to the original feature representations, such as SENET~\cite{huang2019fibinet}, EGate~\cite{huang2020gatenet}, FEN~\cite{yu2019input}, and Dual-FEN~\cite{lu2020dual}. Same as FRNet, we place the above modules after the embedding layer as mentioned in section \ref{sec:pre}. 
In FiBiNET~\cite{huang2019fibinet}, we replace its SENET with other modules to refine the features. Table \ref{Tab:module} shows their performance, and we can make the following conclusions: 
{\em (1) Learning context-aware feature representations is vital for improving the performance of CTR prediction}. Compared with base models, the average improvements (\textit{Avg. Imp}) of FRNet are 0.68\% and 0.80\% in terms of AUC (1.15\% and 17.43\% in terms of Logloss) on Criteo and Frappe, which demonstrates the high effectiveness and compatibility of FRNet. 
{\em (2) FRNet significantly outperforms the other four modules when applied to base models}. FRNet is the only module that can enhance the performance of all seven base models. In contrast, the other four modules may reduce the performance of CTR prediction in some of the datasets or base models. For instance, Applying EGate in FM and AFM achieves poor performance on the Frappe dataset. These phenomena indicate that these model-specific feature refinement modules are with limited compatibility. On the contrary, FRNet has strong compatibility and can be applied in a wide range of CTR prediction models to enhance their performance.

\begin{figure}[t]
{\centering 
\includegraphics[width=0.45\textwidth]{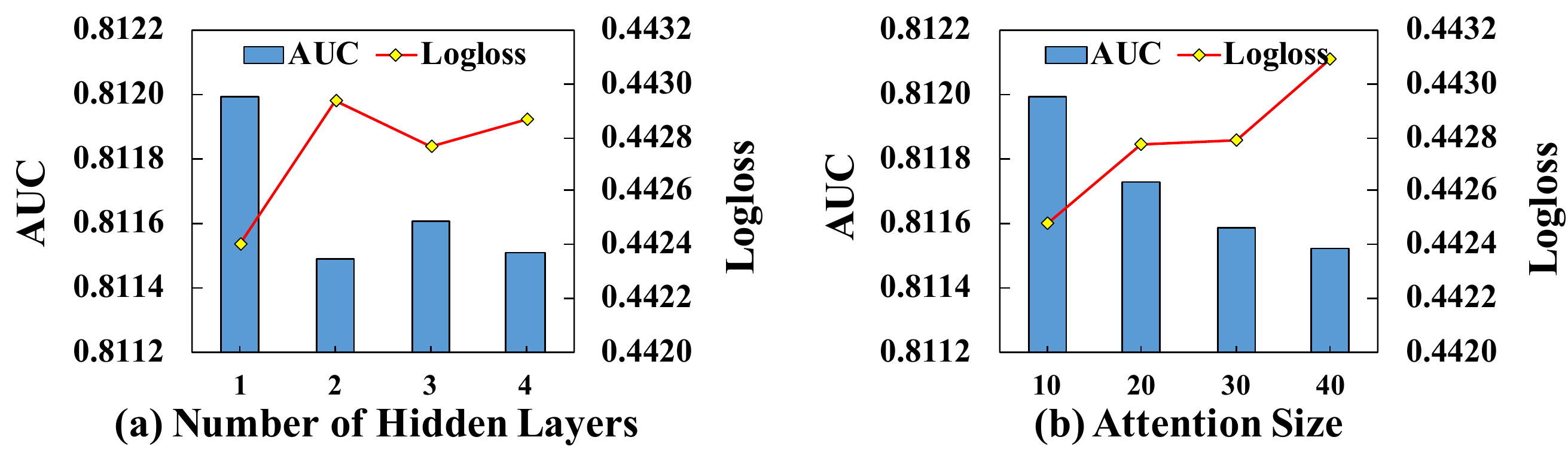}
}\caption{Impact of hyper-parameters on Criteo. }
\label{Fig:hp_criteo} 
\end{figure}

\begin{figure}[t]
{\centering 
\includegraphics[width=0.45\textwidth]{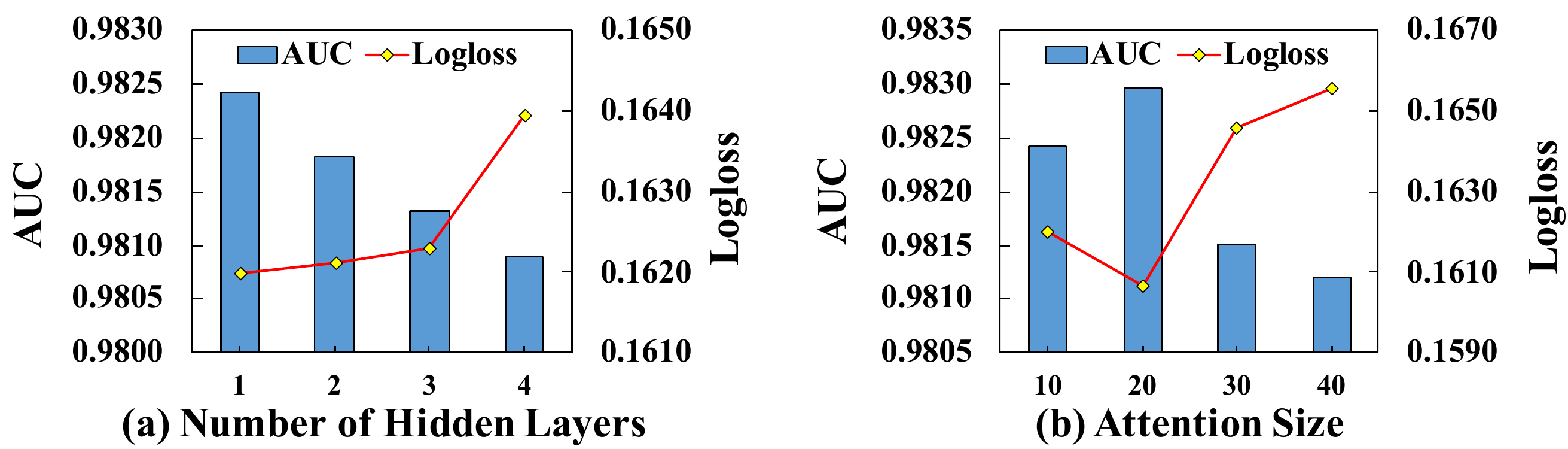}
}\caption{Impact of hyper-parameters on Frappe. }
\label{Fig:hp_frappe} 
\end{figure}

\subsection{Hyper-parameter Study}
\label{sec:hp}
We analyze the impact of hyper-parameters in FRNet, including the number of hidden layers in MLP, the attention size $d_k$ of the Self-Attention. For the sake of convenience, we change the hyper-parameters in $IEU_G$ and $IEU_W$ simultaneously. Note that we only change one hyper-parameter and keep the other one fixed in each experiment.

\textit{Number of Hidden Layers.} Figure~\ref{Fig:hp_criteo} (a) and Figure~\ref{Fig:hp_frappe} (a) show the impact of the number of hidden layers at bit-level unit. For Criteo and Frappe, the most appropriate number of hidden layers is 1. {\em This confirms that the contextual information is not very complicated and a shallow MLP is strong enough to encode contextual information from each instance}.

\textit{Attention Size.} 
As shown in Figure~\ref{Fig:hp_criteo} (b) and Figure~\ref{Fig:hp_frappe} (b), the best attention size for Criteo and Frappe are 10 and 20, respectively. For Criteo, the performance decreases when we increase the attention size. Coincidentally, the dimension of the embedding for Criteo and Frappe are exactly 10 and 20. It may be a good trick to set the attention size to be the same as the embedding dimension.

\begin{figure*}[tb]

\centering
\subfloat[EGate]{
\begin{minipage}[t]{0.19\linewidth}
\centering
\includegraphics[width=1.1\textwidth]{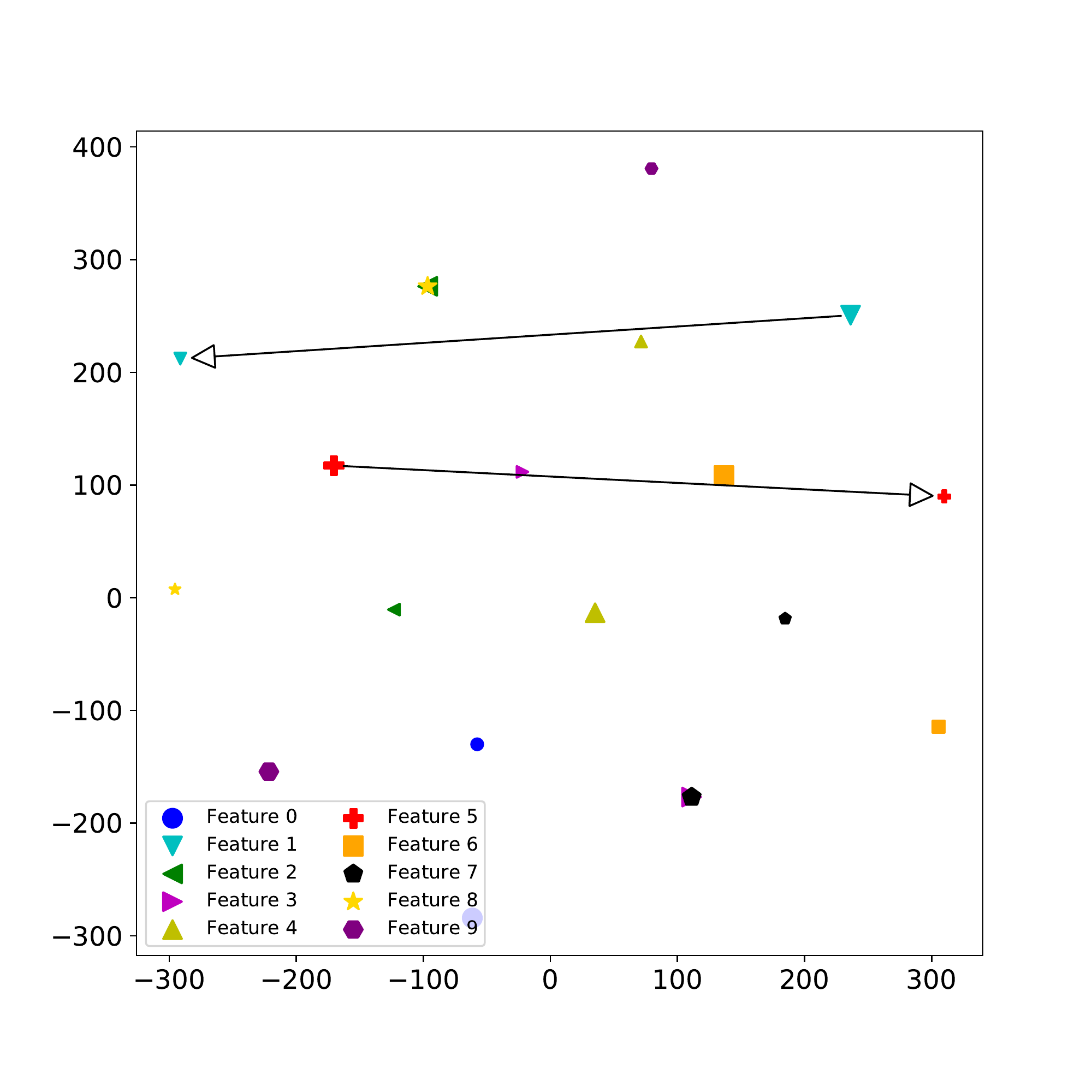}
\end{minipage}
}
\subfloat[DIFM]{
\begin{minipage}[t]{0.19\linewidth}
\centering
\includegraphics[width=1.1\textwidth]{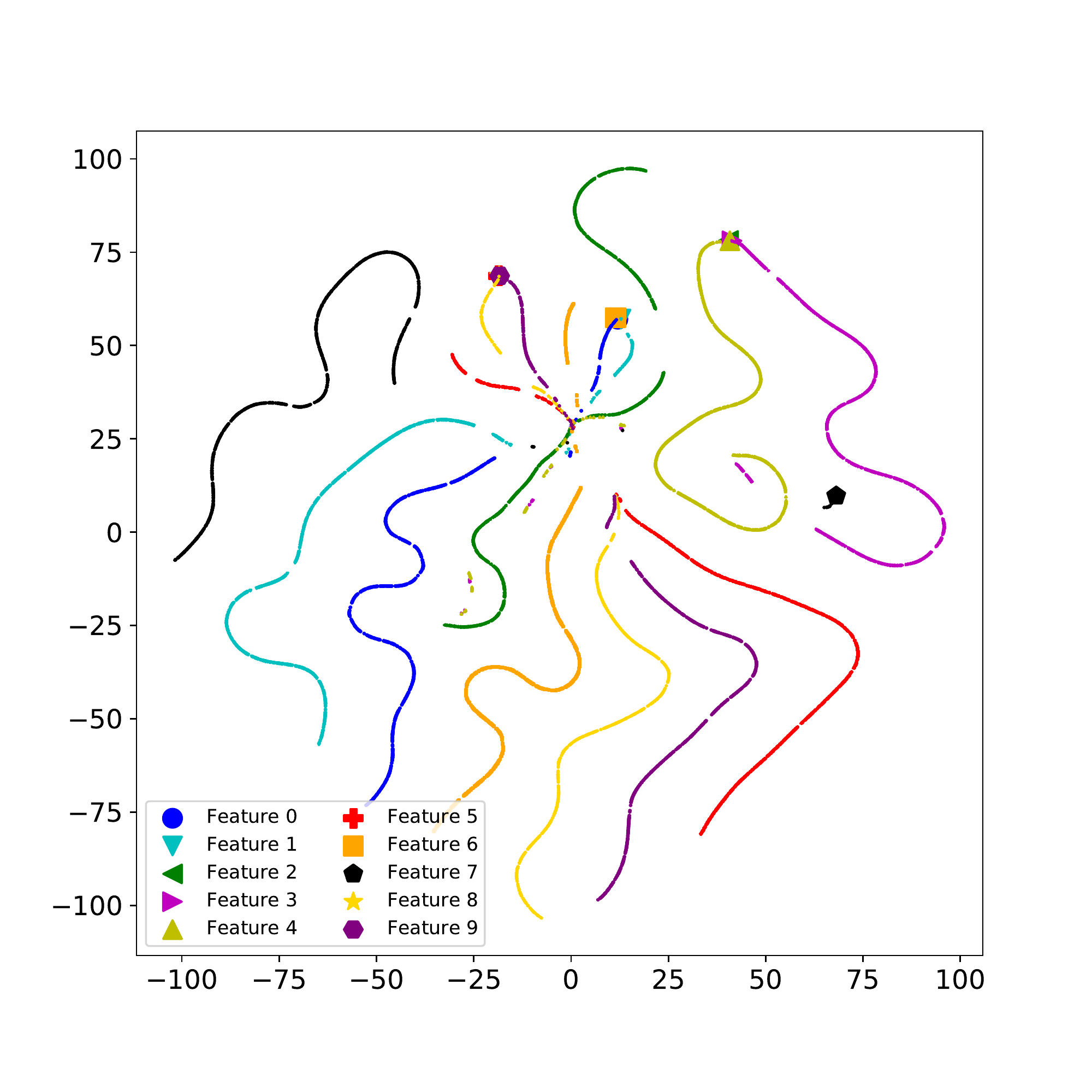}
\end{minipage}
}
\subfloat[Variant \#6]{
\begin{minipage}[t]{0.19\linewidth}
\centering
\includegraphics[width=1.1\textwidth]{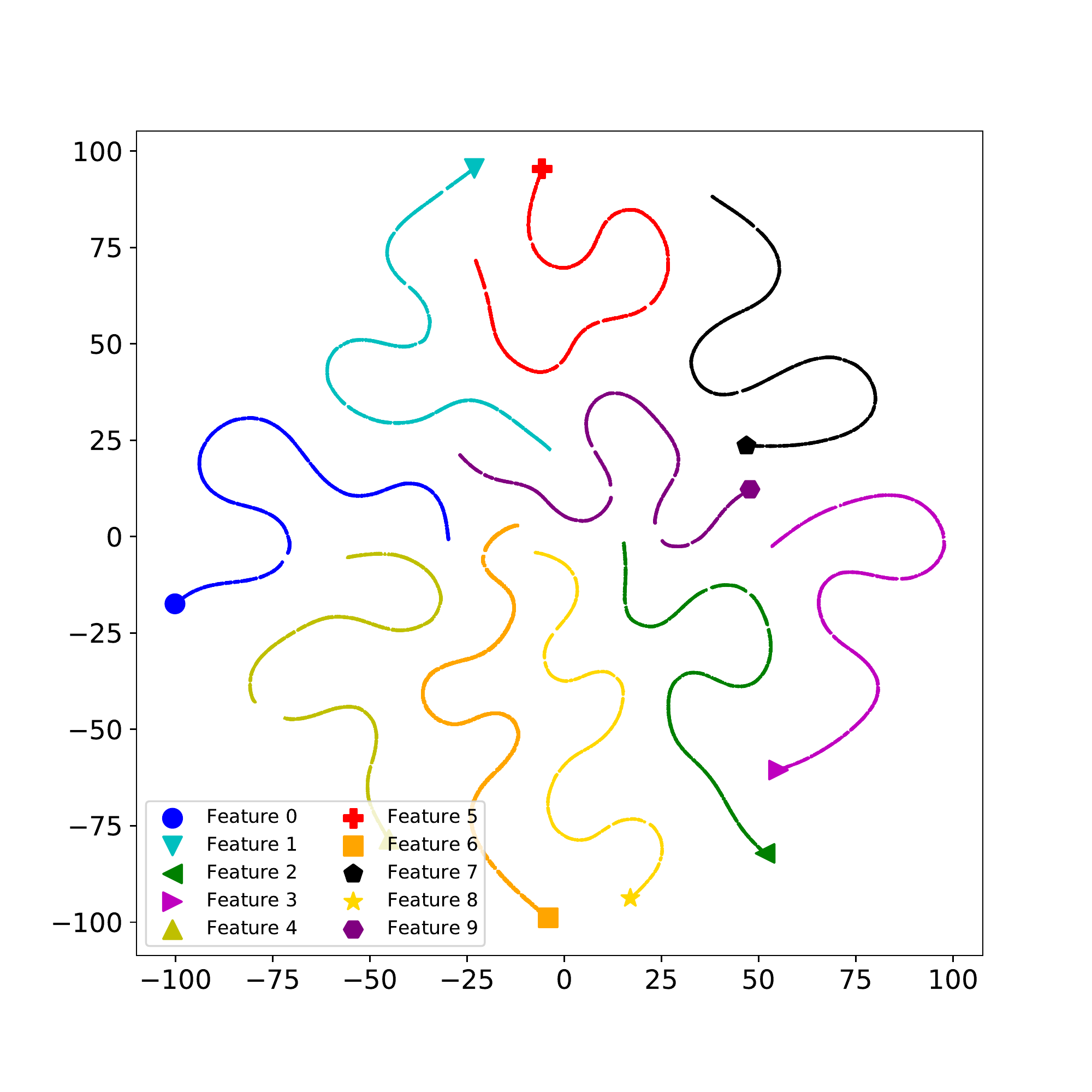}
\end{minipage}
}
\subfloat[FRNet-Vec]{
\begin{minipage}[t]{0.19\linewidth}
\centering
\includegraphics[width=1.1\textwidth]{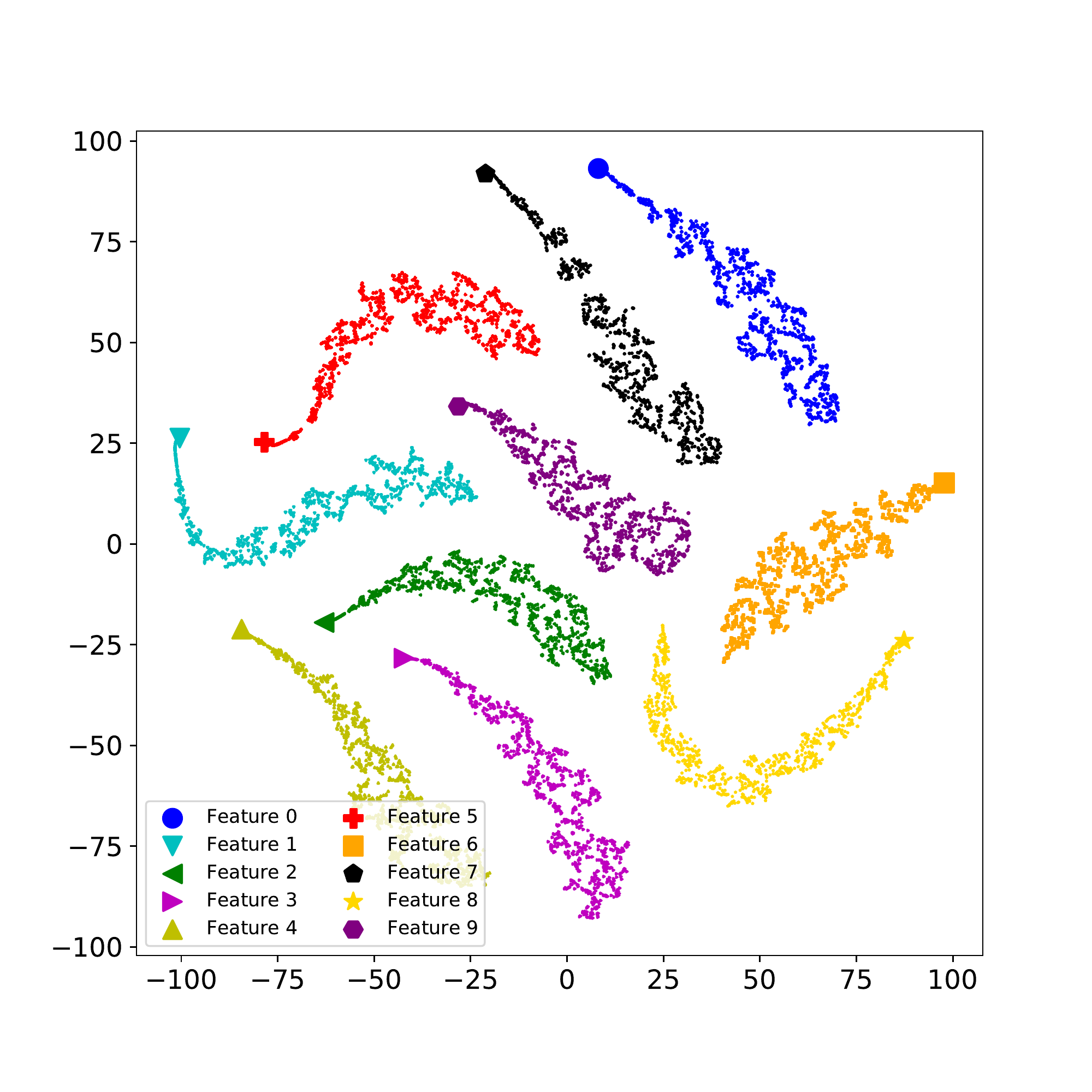}
\end{minipage}
}
\subfloat[FRNet]{
\begin{minipage}[t]{0.19\linewidth}
\centering
\includegraphics[width=1.1\textwidth]{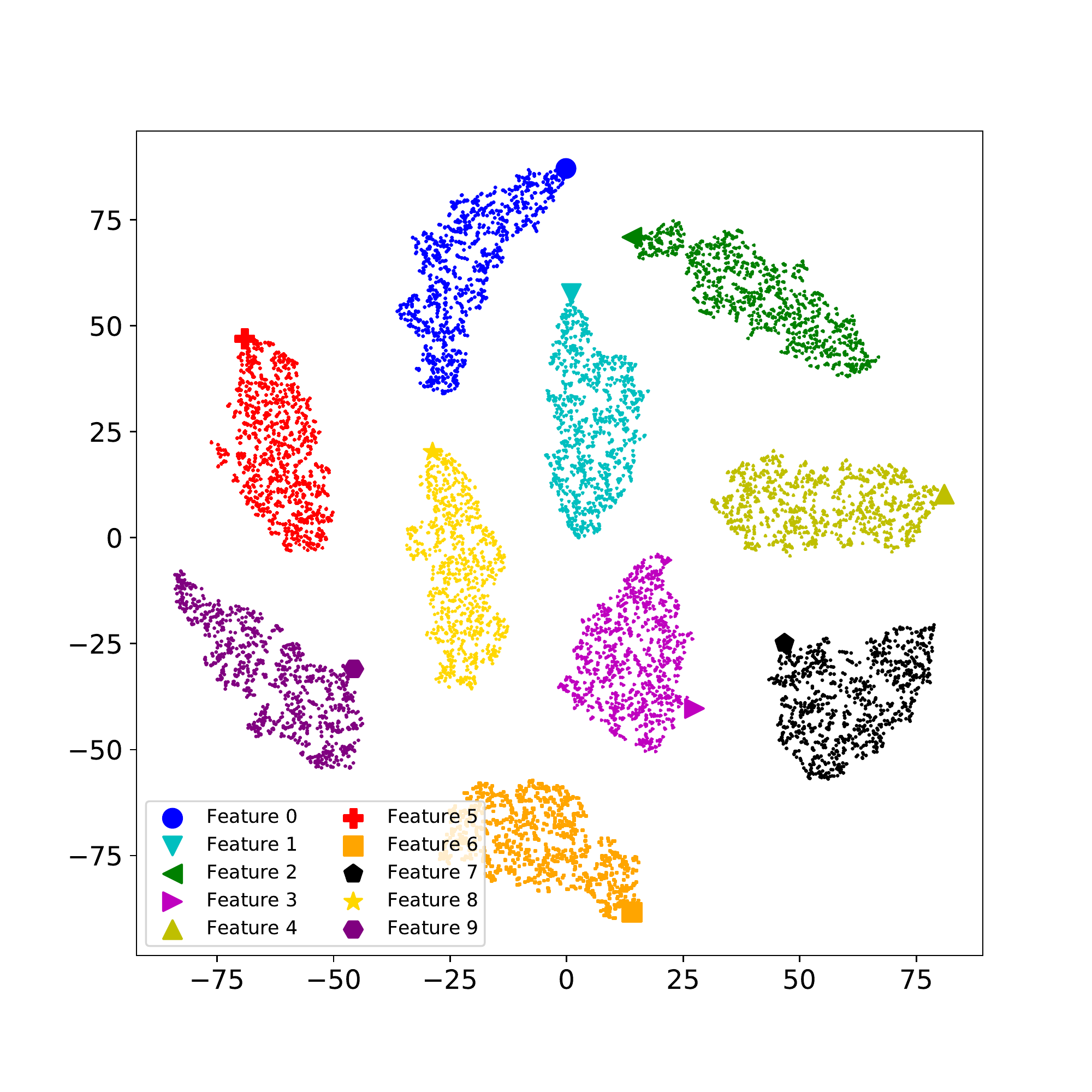}
\end{minipage}
}

\subfloat[Test AUC on Corresponding Subsets of each feature. Each subset contains 1,000 instances.]{
\begin{minipage}[t]{1\linewidth}
\centering
\includegraphics[width=0.9\textwidth]{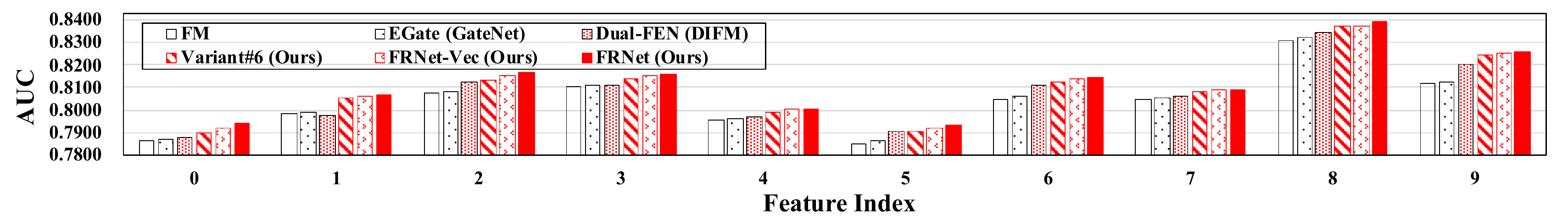}
\end{minipage}%
}
\centering
\caption{Visualization of context-aware feature representations for 10 features in same field. We randomly choose 1000 instances for each feature. Hence, each feature has an original representation and 1,000 context-aware feature representations.}
\label{fig:tsne}
\end{figure*}

\subsection{Ablation Study}
\label{sec:ablation}
Here, we conduct experiments on Criteo and Frappe to prove that each component or design in FRNet plays an essential role in improving the performance of CTR prediction. As shown in Table \ref{tab:abl}, we use the equations to describe how to compute $\mathbf{E}_r$ base on $\mathbf{E}$ by removing or replacing one of the components in FRNet. Especially, variant \#4 denotes that we only use a  self-attention unit in $IEU_G$ and $IEU_W$. 
From Table \ref{tab:abl}, we can make the following conclusions:

\textit{(1) Learning context-aware feature representations is reasonable}. It can be proved that all variants of the FRNet successfully improve the performance of FM on these two datasets; 

\textit{(2) Cross-feature relationships and contextual information are essential.} With cross-feature relationships, variant \#2 outperforms \#1. Meanwhile, \#13 outperforms \#4, and \#3 outperforms \#2, respectively, which shows the effectiveness of contextual information within different instances;

\textit{(3) Assigning weights to original features is valid.} In \#5, we remove $IEU_W$ and then directly add $\mathbf{E}$ and $\mathbf{E_g}$. We can find that \#10 and \#11 outperform \#5, where the learned weights matrix $\mathbf{W_b}$ or $\mathbf{W_v}$ successfully selects important information from $\mathbf{E}$. In addition, \#6 and \#7 outperform \#1, from which we can draw the same conclusion;  

\textit{(4) Learning bit-level weights is more effective than learning vector-level weights.} The variants learning bit-level weights  (\#7, \#9, \#11, \#13) consistently outperform those corresponding variants learning vector-level weights (\#6, \#8, \#10, \#12) respectively, which verifies that learning more fine-grained weights for selecting information is more effective. Intuitively, each element of one feature representation has a specific semantic meaning, so we should give them different weights instead of treating them equally;

\textit{(5) Complementary Features are crucial.}  Variants \#6 and \#7 only learn feature representations from the original feature representations. After adding $\mathbf{E}_g$, \#10 and \#11 outperforms \#6 and \#7 respectively. Furthermore, we observe that \#12 and \#13 outperform \#10 and \#11, as we assign weights $1 - \sigma (\mathbf{W_{b}})$ to $\mathbf{E}_g$, which verifies that assigning weights to complementary features is reasonable. In summary, it is reasonable that the CSGate integrates $\mathbf{E}$ and $\mathbf{E}_g$ with $\mathbf{W_{b}}$. As a comparison, we adopt the idea of Residual Network~\cite{he2016deep} in variants \#8 and \#9 (i.e, adding original representations $\mathbf{E}$), which is also used in DIFM~\cite{lu2020dual}. However, the performance of \#8 and \#9 are worse than \#6 and \#7. The reason is that residual network aims to add the original feature representations to the final feature representations, which might not be enough for CTR prediction.

\begin{table}[t]
\centering
\caption{Ablation study of FRNet. $\mathbf{E}$, $\mathbf{E}_g$ and $\mathbf{E}_r$ denote the original, complementary and context-aware feature representations respectively. $\mathbf{W}_b$ and $\mathbf{W}_v$ denote bit-level and vector-level weights. $\mathbf{O}_{vec}$ is the output of self-attention unit.}
\label{tab:abl}
\scalebox{0.8}{
\begin{tabular}{cccccc} 
\hline\hline
\multirow{2}{*}{Comment/Equation}  & Datasets  & \multicolumn{2}{c}{Criteo} & \multicolumn{2}{c}{Frappe}\\ 
\cline{2-6}
  & Variant & AUC    & Logloss&AUC &Logloss \\
\hline
FM ($\mathbf{E_r} =  \mathbf{E}$)  & \#1 & 0.8028 & 0.4514& 0.9708& 0.1934  \\
\hline
$\mathbf{E_r} = \mathbf{O_{vec}}$                               & \#2 & 0.8056 & 0.4483             & 0.9717& 0.1912\\
$\mathbf{E_r} = \mathbf{E_{g}}$                                 & \#3 & 0.8071 & 0.4470             & 0.9744& 0.1897\\
Removing CIE                                              & \#4& 0.8073 & 0.4468              & 0.9754& 0.1878\\ 
$\mathbf{E_r} =  \mathbf{E} + \mathbf{E_g}$                     & \#5 & 0.8090 & 0.4452             & 0.9778& 0.1821\\

$\mathbf{E_r} =  \mathbf{E} \odot \sigma(\mathbf{W_{v}})$       & \#6 & 0.8110 & 0.4443             & 0.9793& 0.1713\\
$\mathbf{E_r} =  \mathbf{E} \odot \sigma(\mathbf{W_{b}})$       & \#7 & 0.8113 & 0.4437             & 0.9797& 0.1697\\

$\mathbf{E_r} =  \mathbf{E} \odot \sigma(\mathbf{W_{v}}) + \mathbf{E}$    & \#8 & 0.8093 & 0.4452   & 0.9791& 0.1739\\
$\mathbf{E_r} =  \mathbf{E} \odot \sigma(\mathbf{W_{b}}) + \mathbf{E}$    & \#9 & 0.8098 & 0.4449   & 0.9794& 0.1726\\

$\mathbf{E_r} =  \mathbf{E} \odot \sigma(\mathbf{W_{v}}) + \mathbf{E_g}$  & \#10& 0.8110 & 0.4433   & 0.9798& 0.1696\\
$\mathbf{E_r} =  \mathbf{E} \odot \sigma(\mathbf{W_{b}}) + \mathbf{E_g}$  & \#11  & 0.8114 & 0.4430 & 0.9804& 0.1689\\
\hline
FRNet-Vec  & \#12   & 0.8115 & 0.4428   & 0.9816  & 0.1653   \\
\textbf{FRNet}& \#13  & \textbf{0.8120} & \textbf{0.4424 }   & \textbf{0.9830 }   & \textbf{0.1607} \\
\hline\hline
\end{tabular} 
}
\end{table}

\subsection{Visualization of Feature Representations}
\label{sec:vis}

\subsubsection{Visualization Analysis.} To better understand the effectiveness of context-aware feature representations, we first randomly select 10 features from the same field and choose 1,000 instances for each feature from Criteo. Then, we learn the 1,000 feature representations by: (a) EGate, (b) DIFM, (c) Variant\#6, (d) FRNet-Vec and (e) FRNet. Finally, we visualizes their feature representations with t-SNE~\cite{van2008visualizing} in Figure~\ref{fig:tsne}. Each color in Figure~\ref{fig:tsne} represents the original representation of one feature (denoted by the largest symbols, e.g., dots, squares, etc.) and 1,000 different context-aware feature representations in different instances (denoted by smaller symbols).  Variant \#6 is defined in Section  \ref{sec:ablation}, which learns vector-level weights for $\mathbf{E}$ by $IEU_{W}$. Note that we compress the size of feature representations to 2 in this part for the sake of visualization.
As shown in Figure~\ref{fig:tsne}, each feature can learn 1,000 different context-aware feature representations in different instances except EGate, as EGate only produces the fixed feature representations in a specific transformed feature space, where the original feature are mapped to its learned feature 
(as the two arrows show).  From Figure~\ref{fig:tsne}, we have the following observations:

(1) In DIFM, the learned context-aware feature representations among different features are mixed. In contrast, feature representations learned by Variant\#6, FRNet-Vec, and FRNet can be clearly distinguished.

(2) DIFM and Variant \#6 only learn vector-level weights to the fixed original feature representations, so that their refined feature representations should have strictly linear relationships to their original feature representations in high-dimensional feature space. As shown in Figure~\ref{fig:tsne} (b) and (c), the linear relationships are expressed as the continuous curves in the visualization space. However, compared with DIFM, variant \#6 can learn better context-aware feature representations because feature representations are not blended together. Since variant \#6 use IEU to integrate cross-feature relationships and contextual information, this phenomenon confirms that IEU can better distinguish different features. 

(3) FRNet-Vec and FRNet learn nonlinear context-aware features representation for the same feature. FRNet-Vec learns vector-level weights, but after combining with complementary feature representations, the feature representations exhibit strong nonlinear relationship to the original feature representations and the context-aware feature representations for the same feature form a cluster rather than a curve. Different from FRNet-Vec, FRNet simultaneously learns the bit-level weights and complementary features, which further enhances the nonlinearity and the refined feature representations for the same feature form a more diverse cluster. Intuitively, FRNet can learn more different and expressive representations for the same feature under different contexts. 

\subsubsection{Quantitative Analysis.} 

To quantify how the feature representations influence the  performance, we calculate the AUCs of CTR prediction based on the feature representations in Figure~\ref{fig:tsne} (a) - (e) and present the results in Figure~\ref{fig:tsne} (f). DIFM outperforms FM and EGate in most subsets; Variant \#6, FRNet-Vec, and FRNet outperform FM and EGate in all subsets, as FM and EGate only produce fixed feature representation for each feature in different instances. In addition, FRNet learns the most diverse nonlinear context-aware feature representations and achieves the best results than other methods, which further confirms the effectiveness of our method.

\subsection{Visualization of IEU}
\label{sec:attention_analysis}
We design IEU to enable self-attention to incorporate contextual information within different instances. To better understand the effectiveness of IEU, we choose two instances from Criteo with 38 identical features and only one different feature (Feature 0). In the test phase, we input them to $FM_{FRNet}$, and record the output features of $IEU$ and its two components: Self-attention and CIE. 

\begin{figure}
\centering
\subfloat[Self-Attention]{
\begin{minipage}[t]{0.36\linewidth}
\centering
\includegraphics[width=0.90\textwidth]{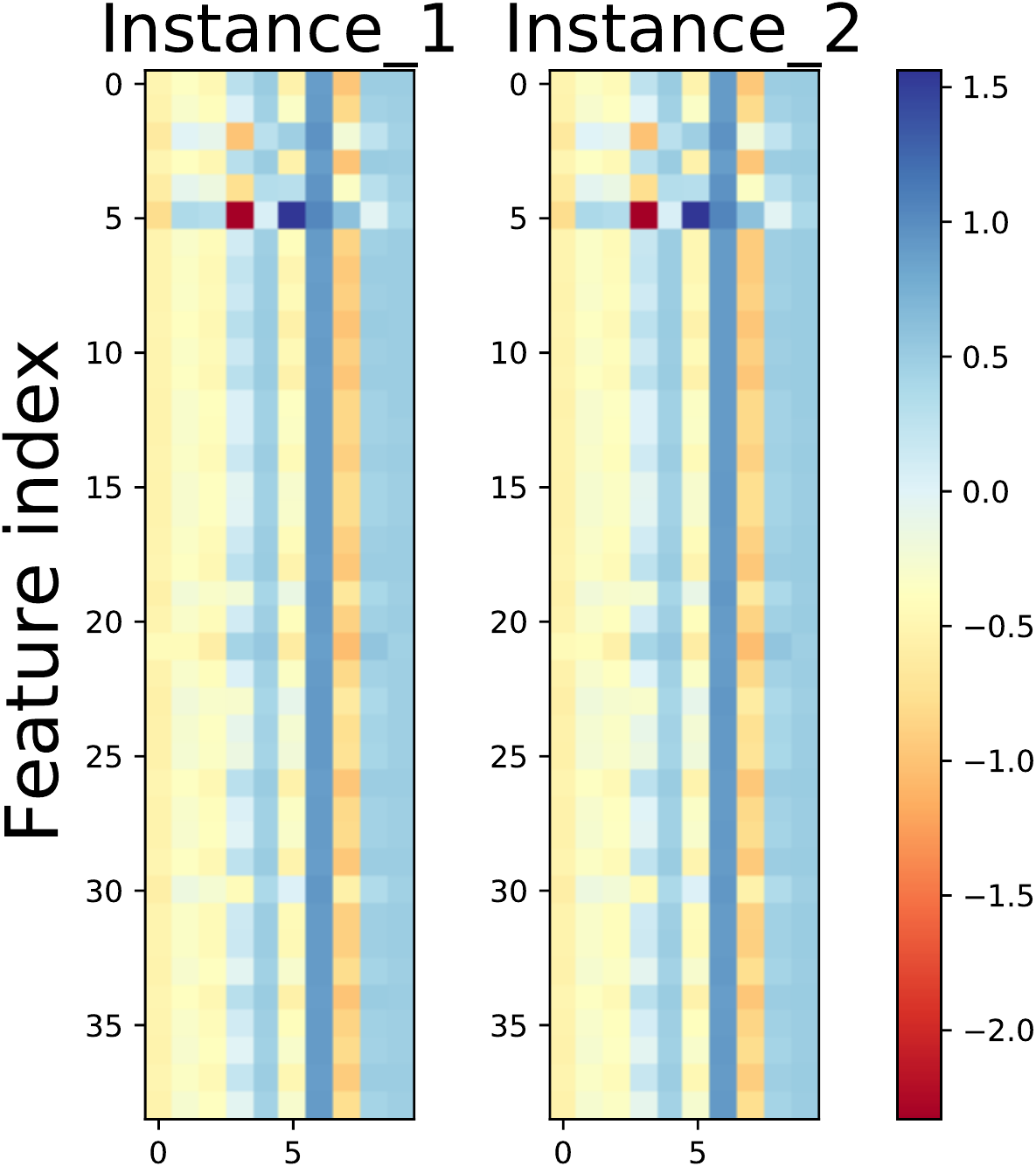}
\end{minipage}
}
\subfloat[CIE]{
\begin{minipage}[t]{0.26\linewidth}
\centering
\includegraphics[width=1.1\textwidth]{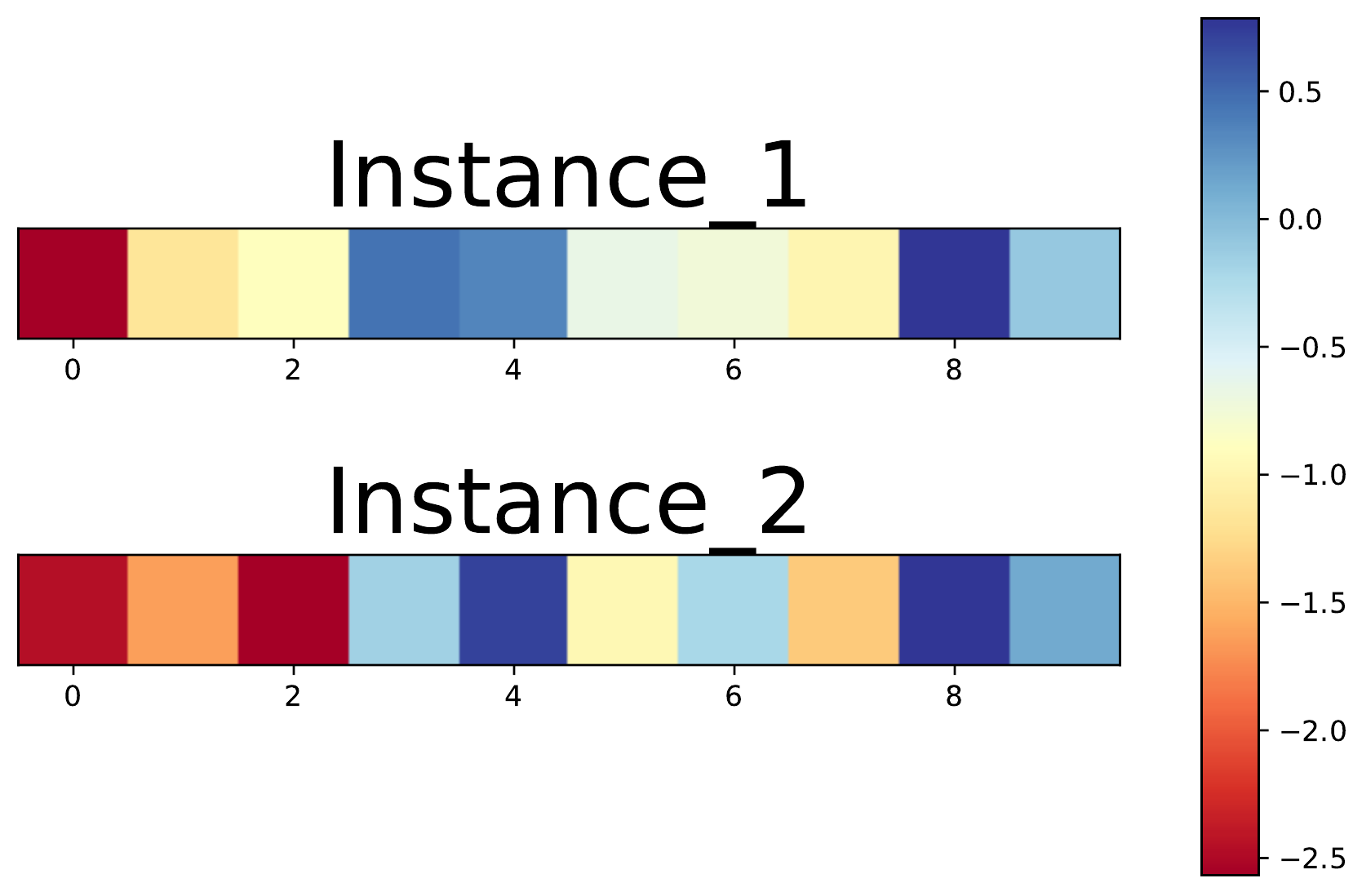}
\end{minipage}
}
\subfloat[IEU]{
\begin{minipage}[t]{0.36\linewidth}
\centering
\includegraphics[width=0.90\textwidth]{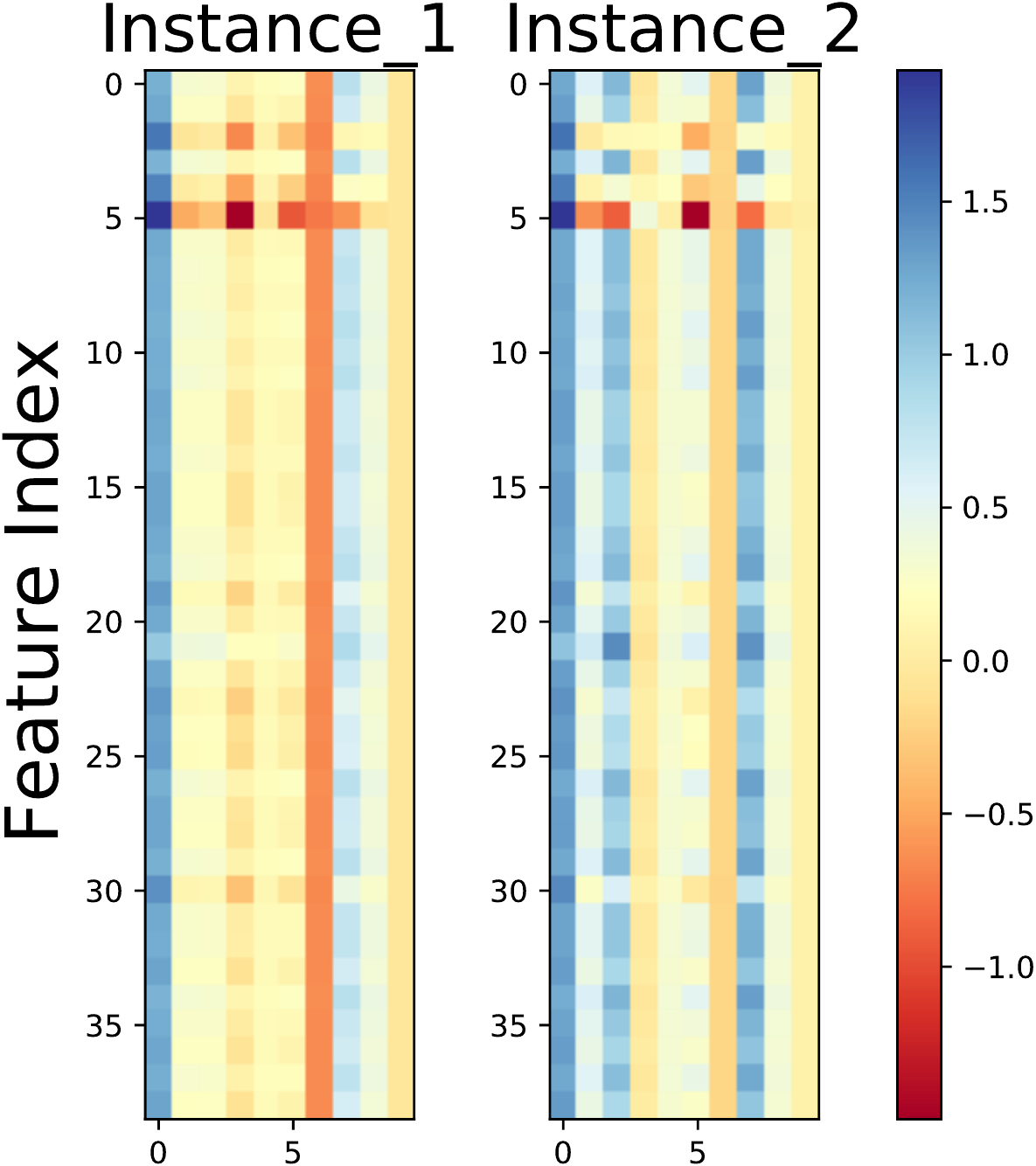}
\end{minipage}
}
\centering

\caption{Heatmap of features learned by IEU and its complements: Self-attention and CIE.}
\label{fig:CIE}
\end{figure}

Figure~\ref{fig:CIE} shows the heatmaps of the features from the three units. As shown in Figure~\ref{fig:CIE} (a), self-attention learns almost identical representations for the same features when the two instances are only with one different feature. Since self-attention only focuses on pair-wise feature interactions in a given instance, it neglects the various contextual information among different instances. In Figure~\ref{fig:CIE} (b), we can see that the two contextual information vectors learned by CIE are with significant differences, which demonstrates that even one different feature can have a significant impact on the two contextual information. Furthermore, $IEU$ integrates the outputs of self-attention and CIE. As shown in Figure~\ref{fig:CIE} (c), for the same feature in the two instances, their representations are significantly different. Furthermore, FRNet utilizes two IEUs, which ensures that it can generate flexible context-aware feature representations for the same feature in different instances.  

\subsection{Distribution of Bit-level Weights}
\begin{figure}
    \centering
    \includegraphics[width=0.25\textwidth]{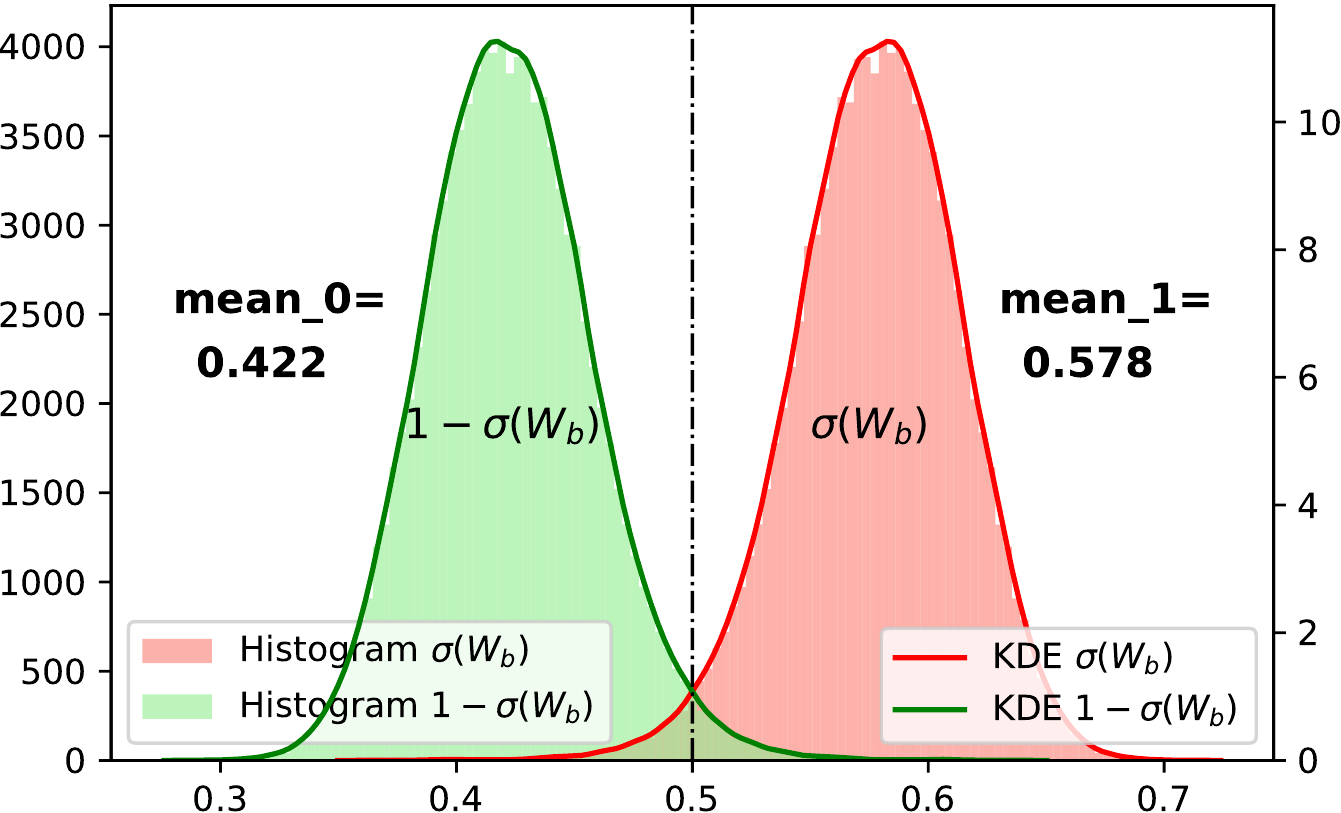}
    \caption{Distribution of bit-level weights in 100K instances.}
    \label{fig:kde}
\end{figure}

Here, we randomly sample 100k instances from the Criteo dataset. We first compute the bit-level weights $\sigma(W_b) \in \mathbb{R}^{39\times10}$ and complementary feature weights $1-\sigma(W_b)$. Then we show the distribution of learned weights (totally 39,000,000 values) in Figure~\ref{fig:kde}. 

We observe that the two distributions follow the normal distribution by observing the histogram and the Kernel Density Estimation (KDE) curve. The values of $\sigma(W_b)$ mean the importance of the original feature representations. On average, the original feature representations are selected by 57.8\%, and the complementary feature representations are selected by 42.2\%. Complementary feature representations boost the performance of FRNet is proved in the ablation study (section \ref{sec:ablation}). This experiment confirms that complementary features are helpful to CTR prediction to a large extent.


\section{Conclusion}
In this paper, we propose a novel module named FRNet, which can learn context-aware feature representations and be used in most CTR prediction models to enhance their performance. In FRNet, we design IEU to integrate contextual information and cross-feature relationships, enabling self-attention to incorporate contextual information within each instance. We also design the CSGate to integrate the original and complementary features representations with learned bit-level weights. Detailed ablation study shows that each design of FRNet contributes to the overall performance. Furthermore, comprehensive experiments verify the effectiveness, efficiency, and compatibility of our proposed method. 

\begin{acks}
This work was supported by the National Natural Science Foundation of China (NSFC) under Grants 61932007 and 62172106.
\end{acks}

\clearpage
\bibliographystyle{ACM-Reference-Format}
\bibliography{frnet_ref}

\end{document}